\documentclass[a4paper,12pt,twoside]{article}
\usepackage{times}
\usepackage[utf8]{inputenc}
\usepackage[english]{babel}
\usepackage[T1]{fontenc}
\usepackage[colorlinks, linkcolor=black, citecolor=black, urlcolor=blue, filecolor=black]{hyperref}
\usepackage{natbib}
\usepackage{graphicx}
\usepackage{subfig}
\usepackage{amssymb,amsmath,amsthm}
\usepackage[top=2cm, bottom=2cm, left=2.5cm, right=2.5cm]{geometry}

\usepackage{lscape}
\usepackage{multirow}
\usepackage{rotating}

\usepackage{setspace}

\usepackage{fancyvrb}
\usepackage{color}
\usepackage{listings}
\definecolor{keywords}{RGB}{128,148,182}
\definecolor{comments}{RGB}{0,0,139}
\title{\Huge{Structural Cohesion: Visualization and Heuristics for Fast Computation}}
\author{\Large{Jordi Torrents}\\\href{mailto:jtorrents@iese.edu}{jordi.t21@gmail.com} \and \Large{Fabrizio Ferraro}\\\href{mailto:fferraro@iese.edu}{fferraro@iese.edu}}
\date{\today}

\begin{document}

\maketitle

\begin{abstract}
The structural cohesion model is a powerful theoretical conception of cohesion in social groups, but its diffusion in empirical literature has been hampered by operationalization and computational problems. In this paper we start from the classic definition of structural cohesion as the minimum number of actors who need to be removed in a network in order to disconnect it, and extend it by using average node connectivity as a finer grained measure of cohesion. We present useful heuristics for computing structural cohesion that allow a speed-up of one order of magnitude over the algorithms currently available. We analyze three large collaboration networks (co-maintenance of Debian packages, co-authorship in Nuclear Theory and High-Energy Theory) and show how our approach can help researchers measure structural cohesion in relatively large networks. We also introduce a novel graphical representation of the structural cohesion analysis to quickly spot differences across networks. 
\end{abstract}


\vspace{2cm}
%\begin{center}
%\begin{large}
%Paper submitted to Social Networks Journal (currently under revise \& resubmit).
%\end{large}
%\end{center}

\vspace{1cm}
\thanks{We, the authors, are in-debted to Aric Hagberg and Dan Schult, developers of the NetworkX python library, for their help in the implementation of the heuristics presented in this paper. We would also like to thank Matteo Prato, Marco Tortoriello, Marco Tonellato, Kaisa Snellman, Francois Collet and Dan McFarland for their comments on early versions of this paper. We gratefully acknowledge funding from the European Research Council under the European Union's Seventh Framework Programme -  ERC-2010-StG 263604 - SRITECH}

\newpage

% \setcounter{tocdepth}{3}
%\tableofcontents
%\newpage
%\doublespacing

Group cohesion is a central concept that has a long and illustrious history in sociology and organization theory, although its precise characterization has remained elusive. Its use in most sociological research has been ambiguous at best. This is largely because, as \citet{moody:2003} argued, it is often based on sloppy operationalization grounded mostly in intuition and common sense. Network analysis has provided a large number of solutions to this problem. From classical work in the graph-theoretic sociological tradition on cliques, clans, clubs, $k$-plexes, $k$-cores and lambda sets \citep[chapter 8]{wasserman:1994}, to the more recent contribution of physicists and computer scientists on community analysis \citep{fortunato:2010}, network theorists have provided researchers with a wide range of measures of cohesion in social networks.

However, neither the classical approaches nor new developments in community analysis are well-enough suited to address many of the common uses of group cohesion in the sociological and organizational literature, for three key reasons. First, while most of these measures can help us identify cohesive subgroups, they do not provide insight into their robustness, which is a critical element to the theoretical conceptualization of cohesion. In most cases, the removal of only a few actors from the subgroups can lead to its fragmentation into smaller disconnected groups \citep{white:2001}. Secondly, many cohesive subgroup measures do not allow for overlap among subgroups. Finally, even when they do allow for overlap, most measures cannot capture the hierarchical nature of nested social groups, where subgroups, like Russian dolls, are recursively nested in one another. As a result, hardly any of the existing measures capture the theoretical complexity of cohesion, and thus fall short of offering useful operationalizations for many empirical phenomena of sociological interest.

One model which provides a more fertile ground for sociological analysis is the structural cohesion model \citep{white:2001,moody:2003}. This model is grounded on two common conceptualizations of group cohesion in the literature. A social group is considered cohesive to the extent that: a) it is resistant to being pulled apart by the removal of some of its members; and b) pairs of its members have multiple direct or indirect connections that pull it together \citep[309-310]{white:2001}. Building on the concept of node connectivity from graph theory, the structural cohesion of a group is defined in this model as the \emph{minimal number of actors who need to be removed from the group to disconnect it}. Despite its solid and elegant mathematical foundation, the structural cohesion model has not been widely used in empirical analysis because it is not possible to perform the required computations for networks with more than a few thousands nodes and edges in a reasonable time frame.

These computational challenges also hindered the development of an interesting feature of the structural cohesion model: its applicability to both bipartite and unipartite networks. While many social networks are essentially bipartite in nature (as people meet, interact, and collaborate around specific events and/or objects), most of our analytical tool-kit was developed to analyze one-mode networks \citep*{latapy:2008}. Therefore it was common practice to conduct network analysis on one-mode projections only, but it is now clear that this practice leads to biased estimates of key measures, as recent work on the clustering coefficient has amply shown \citep{robins:2004,lind:2005,latapy:2008}. The structural cohesion model, instead, can be applied without modification to both bipartite and unipartite networks \citep*{white:2004}. That said, the original algorithm is prohibitively time-consuming to compute, especially with the exponential growth in the size of available network data. 

In this paper we extend the structural cohesion model by using the concept of average node connectivity, that is the \emph{average number of actors who need to be removed from the group to disconnect an arbitrary pair of actors in  the group}. We present a set of heuristics to compute structural cohesion based on the fast approximation to compute pairwise node independent paths \citep{white:2001b}. We implemented it in NetworkX \citep{hagberg:2008}, a Python Library for Complex Network Analysis. The heuristics presented here allow us to compute the approximate value of group cohesion for moderately large networks, along with all the hierarchical structures of connectivity levels, one order of magnitude faster than implementations which are currently available. We also suggest a novel graphical representation of the results of the analysis that might help synthetically communicate results and spot differences across different networks \citep*{moody:2005}.

We used our implementation of the heuristics proposed in this paper to analyze three large collaboration networks: the co-maintenance network of Debian packages, and the co-authorship networks in Nuclear Theory and High-Energy Theory. We ran our analysis in both one-mode and two-mode networks, and compare the networks in terms of their connectivity structure. Consistent with the literature on two-mode networks, we show that the complex hierarchy of collaboration captured in the two-mode analysis is a better representation of the connectivity structure of empirical networks than their one-mode counterparts.

The rest of the paper is organized as follows: we start by laying out the notation we use in the rest of the paper. Then we discuss the main features which a cohesive subgroup formalization should have from a sociological perspective, reviewing the most important formalizations of cohesive subgroups in the social network literature and discussing in depth the structural cohesion model. We then describe the exact algorithm proposed by \citet{moody:2003} to compute the connectivity hierarchy of a given network. After that, we introduce our proposed heuristics, and describe their implementation and performance. We go on to report our findings from applying the structural cohesion analysis to three large collaboration networks, as well as proposing a novel graphical representation of the connectivity structure using a three-dimensional scatter plot. Finally we conclude with implications for future research.

\section{Terminology and notation}

An undirected graph $G=(V,E)$ consists of a set $V(G)$ of $n$ nodes and a set $E(G)$ of $m$ edges, each one linking a pair of nodes. The \emph{order} of $G$ is its number of nodes $n$ and the \emph{size} of $G$ is its number of edges $m$. Two nodes are adjacent if there is an edge that links them, and this edge is said to be incident with the two nodes it links. A \emph{subgraph} of $G$ is a graph whose nodes and edges are all in $G$. An \emph{induced subgraph} $G[U]$ is a subgraph defined by a subset of nodes $U \subseteq V(G)$ with all the edges in $G$ that link nodes in $U$. A subgraph is \emph{maximal} in respect to some property if the addition of more nodes to the subgraph will cause the loss of that property.

A \emph{path} is an alternating sequence of distinct nodes and edges in which each edge is incident with its preceding and following nodes. The length of a path is the number of edges it contains. The \emph{shortest path} between two nodes is a path with the minimum number of edges. The \emph{distance} between any two nodes $u$ and $v$ of $G$, denoted $d_{G}(u,v)$, is the length of the shortest path between them. The \emph{diameter} of a graph $G$, denoted $diam(G)$, is the length of the longest shortest path between any pair of nodes of $G$. \emph{Node independent paths} are paths between two nodes that share no nodes in common other than their starting and ending nodes. A graph is connected if every pair of nodes is joined at least by one path. A \emph{component} of a graph $G$ is a maximal connected subgraph, which means that there is at least one path between any two nodes in that subgraph.

The \emph{density} of a graph $G$, denoted $\varrho(G)$, measures how many edges are in set $E(G)$ compared to the maximum possible number of edges among nodes in $V(G)$. Thus, density is calculated as $\varrho(G) = \frac{2m}{n(n-1)}$. A \emph{complete graph} is a graph in which all possible edges are present, so its density is 1. A \emph{clique} is an induced subgraph $G[U]$ formed by a subset of nodes $U \subseteq V(G)$ if, and only if, the induced subgraph $G[U]$ is a complete graph. Thus, there is an edge that links each pair of nodes in a clique. The \emph{degree} of a node $v$, denoted $deg(v)$, is the number of edges that are incident with $v$. The minimum degree of a graph $G$ is denoted $\delta(G)$ and it is the smallest degree of a node in $G$. A $k$-core of $G$ is a maximal subgraph in which all nodes have degree greater or equal than $k$; which means that a $k$-core is a maximal subgraph with the property $\delta \ge k$. The \emph{core number} of a node is the largest value $k$ of a $k$-core containing that node.

The removal of a node $v$ from $G$ results in a subgraph $G - v$ that does not contain $v$ nor any of its incident edges. The \emph{node connectivity} of a graph $G$ is denoted $\kappa(G)$ and is defined as the minimum number of nodes that must be removed in order to disconnect the graph $G$. Those nodes that must be removed to disconnect $G$ form a \emph{node cut-set}. If it is only necessary to remove one node to disconnect $G$, this node is called an \emph{articulation point}. We can also define the \emph{local node connectivity} for two nodes $u$ and $v$, denoted $\kappa_{G}(u,v)$, as the minimum number of nodes that must be removed in order to destroy all paths that join $u$ and $v$ in $G$. Then the \emph{node connectivity} of $G$ is equal to $min{\{\kappa_{G}(u,v):u,v \in V(G)\}}$. Similarly, the \emph{edge connectivity} of a graph $G$ is denoted $\lambda(G)$ and is defined as the minimum number of edges that must be removed in order to disconnect the graph $G$. The edges that must be removed to disconnect $G$ form an \emph{edge cut-set}.

The measures discussed above are defined as properties of whole graphs but they can also be applied to subgraphs. A \emph{$k$-component} is a maximal subgraph of a graph $G$ that has, at least, node connectivity $k$: we need to remove at least $k$ nodes to break it into more components. The \emph{component number} of a node is the largest value $k$ of a $k$-component containing that node. Notice that $k$-components have an inherent hierarchical structure because they are nested in terms of connectivity: a connected graph can contain several 2-components, each of which can contain one or more tricomponents, and so forth. 

\section{Cohesion in social networks}

\citet{doreian:1998} argue that group cohesion can be divided analytically into an \emph{ideational} component, which is based on the members' identification with a collectivity, and a \emph{relational} component, which is based on connections among members. These connections are, at least in part, observable, and thus the relational approach seems more appropriate for theory building and empirical research. But, despite its attractiveness, the relational component has received much less attention than the ideational component in sociological literature. Social network analysis has been the exception, and since the beginning, its proponents formalized group cohesion in relational terms, that is, they defined the boundaries of subgroups in a community starting from the patterns of relations among actors.

Unfortunately most of the existing formalizations of cohesive subgroups do not capture some key properties of the theoretical concept of cohesive groups. First, a cohesive subgroup should be \emph{robust}, in the sense that its qualification as a group should not be dependent on the actions of a single individual, or any small set of individuals that belong to the group. This implies, on the one hand, that no actor, or small set of actors, should be able to dissolve the cohesive subgroup by abandoning it; while, on the other hand, all actors in a group should be related to all other actors by multiple direct or indirect connections in order to pull it together \citep{white:2001,moody:2003}. Therefore, cohesive subgroups should also be relatively invariant to changes outside the group \citep[chapter 6]{brandes:2005}.

Second, actual social groups tend to  \emph{overlap} in the sense that some actors are likely to be part of more than one cohesive subgroup. As \citet{freeman:1992} notes, formalizations of subgroups that overlap a lot are not well suited to capturing the theoretical concept of groups because their sociological use is not focused on individuals but on contexts, such as productive relations, friendship relations, or family ties, to name a few. Thus if groups are defined around a higly specific context the overlap is likely to be small. Therefore the formalization of subgroups often assumed non-overlapping subgroups. Moreover, non-overlapping subgroups can be used to develop categorical variables for membership that could be used in regression analysis \citep{borgatti:1990}. However, there is always overlap among cohesive subgroups in actual social groups; and this overlap might be both empirically and theoretically relevant.

Third, following a typical distinction in the social network literature, cohesive groups have both a \emph{structural} and a \emph{positional} dimension. In the former, cohesive subgroups are defined in terms of the global patterns of relations, and the focus is on the groups and the network as a whole. In the latter, the focus is on the identification of actors who, because of their network position, obtain preferential access to information or resources that flow through the network. Cohesive subgroup formalizations should help address both structural and positional questions. 

Last but by no means least, cohesive subgroups are likely to display a \emph{hierarchical structure} in the sense that highly cohesive subgroups are nested inside less cohesive ones. This notion of hierarchy is grounded on Simon's definition: ``a system that is composed of interrelated subsystems, each of the latter being, in turn, hierarchic in structure until we reach some lowest level of elementary subsystem'' \citep[468]{simon:1962}. A hierarchical conception of cohesive subgroups implies that there is a relevant organization at all scales of the network, and that cohesive groups are a mesolevel structure that is not reducible to neither macro nor micro level phenomena and dynamics. This nested conception of cohesive subgroups provides a direct link with the structural dimension of the sociological concept of embeddedness \citep{granovetter:1985}. The nested nature of cohesive groups allows one to operationalize social relations that are, in direct contrast to arms length relations, structurally embedded in a social network.

In the following section we briefly review existing social network formalizations of subgroup cohesion.  For each method, in table \ref{t:cohesive} we provide the definition, the underlying logic, the measure proposed, and evaluate them in terms of the four criteria just described. We will therefore consider whether they are robust, can allow for overlapping groups, provide information on both the structure and the position of nodes, and whether they capturethe hierarchical structure of the groups.

\subsection{Formalizations of cohesive subgroups}

Historically, the first social networks approaches to subgroup cohesion formalization identified cohesive subgroups by considering only internal ties among the actors in the group. However, most recent formalizations define cohesive subgroups by considering both internal ties among its members and also external ties between each subgroup and the rest of the network \citep{wasserman:1994}. All the formalizations based on internal ties are based on the concept of clique, which were later generalized by relaxing some of the strict conditions of distance, degree or density that the clique concept imposes.  The formalizations that consider both internal and external ties can be organized in two main categories depending on whether they use density or connectivity to measure internal and external ties.

The first formalization of cohesive subgroups was the concept of clique \citep{luce:1949}, which is a maximal subset of actors in which each actor is directly connected to every other actor in the subgroup. For small groups in some contexts, such as friendship networks, it makes sense to use the clique concept. However, in many contexts, especially in large and/or very sparse networks, it is unlikely that the existing cohesive subgroups will be formed by actors that have direct relations with all other actors in the subgroup. Cliques, however, intuitively capture the idea that a cohesive subgroup exists independently of the action of any individual in the group. Thus the group is robust because it cannot be disconnected by removing any individual actor. Cliques can overlap ---and they usually do so a lot--- but they do not display a hierarchical organization. Because of the limitations of the clique concept, some generalizations were developed; on the one hand, there emerged a family of generalizations based on relaxing distances among members of the subgroup ---$n$-cliques , $n$-clans, and $n$-clubs \citep{mokken:1979}; and, on the other, generalizations based on relaxing the number of links between members of the subgroup ---$k$-plex \citep{seidman:1978}, and $k$-cores \citep{seidman:1983}.

All these generalizations except for $k$-core are quite arbitrary because the analyst has to set the parameters $n$ or $k$ depending on the concrete aim of the analysis at hand and its empirical setting. Thus, $k$-core is the only generalizationof the clique concept with an inherent hierarchical structure: 3-cores are always nested inside 2-cores; and 4-cores inside 3-cores, and so forth. Thus, this formalization captures an important aspect of the sociological concept of cohesive groups. However, $k$-cores are not robust because the removal of a few actors could potentially disconnect them; in fact they don't even need to be connected at all to be a $k$-core \citep{white:2001}. Furthermore, the definition of $k$-core only considers internal relations among actors within it, without considering relations with the rest of the network.

Another important subset of subgroup formalizations identifies cohesive subgroups by comparing the internal and external ties of subgroups members. The two key criteria to define groups in these categories are density and connectivity. The first formalization of this kind was the LS set \citep{luccio:1969,lawler:1973}: a set of nodes in which each of its proper subsets has more ties with the nodes outside that subset than the LS set itself. The main idea is that an LS set is a union of subsets of nodes. This union is better than any subset in terms of cohesion because it has fewer connections to the outside. Thus, actors in the LS set have more connections to other members than to outsiders. LS sets are robust to the removal of edges and they have an inherent hierarchical structure; however, due to their strict requirements, only very few LS sets are actually found in empirical social networks. Lambda sets \citep{borgatti:1990} were introduced as a generalization of LS sets designed to capture only the edge-connectivity properties of the LS sets. Lambda sets are maximal subsets of nodes that have more edge independent paths between them than with nodes outside the subset. This generalization, however, does not capture important features of the sociological concept of group cohesiveness. On the one hand, they are not robust to the removal of nodes, and, on the other hand, the edge independent paths that link the members of a Lambda set can go through nodes that are not in the lambda set, thus there is no strict separation between the role of actors inside and outside a lambda set in respect to its internal cohesion. 

\begin{landscape}

\begin{table}
\begin{footnotesize}
\begin{tabular}{|c|p{1.7cm}|p{2.7cm}|p{2cm}|p{6cm}|p{1.7cm}|p{1.7cm}|p{1.7cm}|p{1.7cm}|p{2.7cm}|}
\hline
&Based on&Criteria&Measure&Definition&Robust&Overlap&Positional&Hierarchical&Computational\\
\hline
% Local measures
\multirow{6}{*}{\begin{sideways}\mbox{Absolute: only internal}\end{sideways}}&
    % Clique
    complete\newline connectivity&$diam=\varrho=1$\newline$\delta=\lambda=\kappa=n-1$&clique
    &maximal subgraph of nodes all of which are adjacent to each other
    &Yes&Yes: clique percolation&Yes: structural folds&Yes:\newline$k$-cliques&Slow\\
    \cline{2-10}
    % Relax distance
    &\multirow{3}{*}{relax distance}
        % N-clique
        & $max{\{d_{G}(u,v)\}} \le n $&$n$-clique&
        maximal subgraph in which the largest geodesic distance is no greater than $n$
        &No&No&No&No&Slow\\
        \cline{3-10}
        % N-clan
        &&$n$-clique with\newline $diam \le n$&$n$-clan&
        $n$-clique that also have a diameter no greater than $n$&No&Yes&No&No&Slow\\
        \cline{3-10}
        % N- club
        &&$diam=n$&$n$-club&a maximal subgraph of diameter $n$&No&Yes&No&No&Slow\\
        \cline{2-10}
    % Relax degree
    &\multirow{2}{*}{relax degree}
        % K-plex
        &$\delta \ge n - k$&$k$-plex&maximal subgraph in which each node may be lacking ties to
        no more than $k$ other nodes&No&Yes&No&No&Slow\\
        \cline{3-10}
        % K- core
        &&$\delta \ge k$&$k$-core&maximal subgraph in which all nodes have degree $k$ or more
        &No&No&No&Yes&Very fast $O(m)$\\
        \cline{2-10}
    % Relax density
    &relax density& $\varrho \ge \eta $&$\eta$-dense subgraph
    &subgraph with density greater than or equal to $\eta$, where $0 \le \eta \le 1$
    &No&No&No&No&Slow\\
\hline
\hline
% Global measures
\multirow{5}{*}{\begin{sideways}\mbox{Relative: Internal (+) External (-)}\end{sideways}}&
    % Density based
    \multirow{2}{*}{density}
        % LS sets
        &minimize edges to\newline outside&LS sets&set of nodes in which each of its proper subsets
        has more ties with the nodes outside that subset than the LS set itself
        &Yes&No&Yes&Yes&Slow $O(n^4)$\\
        \cline{3-10}
        % community algorithms: modularity
        &&quality function of\newline partitions&modularity&the fraction of the edges that fall within
        the given groups minus the expected such fraction if edges were distributed at random
        &No&No&No&No&Optimum: Slow\newline Approx: Fast\\
        \cline{2-2} \cline{4-5}
    % Connectivity based
    &\multirow{3}{*}{connectivity}
        % Community algorithms: conductance
        &&conductance&weight of edge cut-sets among different subgroups&&&&& \\
        \cline{3-10}
        % Lambda sets
        &&edge-connectivity&lambda sets&maximal subset of nodes that have more edge
        independent paths between them than with nodes outside the subset&Not as robust as LS sets&No&No&Yes
        &Slow $O(n^4)$\\
        \cline{3-10}
        % K-components
        &&node-connectivity&$k$-components&maximal subgraph that has, at least, node connectivity $k$:
        we need to remove at least $k$ nodes to break it into more components&Yes&Yes: $k-1$ nodes&Yes
        &Yes&Exact: Slow $O(n^4)$\newline Approx: $\ll O(n^4)$\\
\hline
\hline
% Other measures?
\multicolumn{5}{|c|}{random walk based partition algorithms}&No&No&No&No&Fast\\
\hline
\end{tabular}
\end{footnotesize}
\caption{Summary of cohesive subgroups formalizations from social network analysis literature \citep{luce:1949,luccio:1969,lawler:1973,seidman:1978,mokken:1979,seidman:1983,seidman:1983ls,borgatti:1990,wasserman:1994,white:2001,moody:2003,brandes:2005,fortunato:2010}. Notation: $diam$ is diameter, $\varrho$ is density, $\delta$ is minimum degree, $\lambda$ is edge-connectivity, $\kappa$ is node connectivity, $n$ is the number of nodes, $m$ is the number of edges, and $d_{G}(u,v)$ is the distance between nodes $u$ and $v$ in $G$.}
\label{t:cohesive}
\end{table}

\end{landscape}

More recently, under the label community analysis, an interdisciplinary community of researchers interested in complex networks has proposed a novel family of subgroup measures and algorithms \citep{fortunato:2010}. Essentially their approach is to divide a network into subgroups by grouping nodes that are more densely connected among them than with the rest of the network. To objectively define how good a concrete partition of a network is, they define a quality function \citep{brandes:2005,fortunato:2010}. There are many different quality functions used in network literature, with most of them based on density, but also a few based on connectivity. The most popular quality function is modularity, which is computed as the fraction of the edges that fall within the given groups minus the expected value of the fraction if edges were distributed at random. However, the subgroups resulting from community analysis techniques are not hierarchically organized in the sociological sense discussed above because there is no natural nestedness among groups\footnote{However, some of those methods are called hierarchical because they use hierarchical clustering to organize partitions in each step of the partition algorithm, which is commonly represented by a dendogram. Thus, researchers need to to introduce an arbitrary criteria to identify relevant partitions --that is, the level at which we cut the dendogram.}.

The first wave of community analysis focused on the analysis of non overlapping groups, but recent developments have explored overlapping community structures. The most interesting approach of this kind is the clique percolation method \citep*{palla:2005} and their generalizations based on short cycles connectivity \citep{batagelj:2007}. A $k$-clique is a complete subgraph formed by $k$ members. Two $k$-cliques are considered adjacent if they share $k-1$ actors. A $k$-clique community is the largest connected subgraph obtained by the union of all adjacent $k$-cliques. $k$-clique communities can share nodes, so overlapping is possible. The clique percolation approach has proven to be a fertile ground over which to build theoretical developments on the positional dimension of cohesion. The concept of intercohesion based on the structural fold network topology \citep{stark:2010} is the most prominent example. Actors at structural folds are insiders in multiple cohesive subgroups ($k$-clique communities). Thus they have access to diverse resources and information from each subgroup without being isolated and limited to only one group of neighbors. \citeauthor{stark:2010} show that this distinctive structural position helps to explain innovation and entrepreneurial dynamics in the context of firm networks.

However these new developments on community analysis are not well suited to address many of the common uses of group cohesion in the sociological literature. The clique percolation method assumes that the network under analysis has a large number of cliques, so it may fail to deliver meaningful results for networks with few cliques; also, if there are too many cliques, it may yield trivial results, such as considering the whole network a cohesive group without internal divisions. Moreover, this method is focused on finding subgraphs that contain many $k$-cliques inside, which is not exactly the same as subgraphs more densely connected internally than externally, because a $k$-clique community could be formed by chains of $k$-cliques with low edge density among non adjacent $k$-cliques. This implies that $k$-clique communities are not necessary robust to node removal. 

\subsection{The structural cohesion model}

The structural cohesion approach to subgroup cohesion \citep{white:2001,moody:2003} is grounded on two mathematically equivalent definitions of cohesion that are based on commonly used concepts of cohesion in the sociological literature. On the one hand, the ability of a collectivity to hold together independently of the will of any individual. As set out by the formal definition, ``a group's structural cohesion is equal to the minimum number of actors who, if removed from the group, would disconnect the group'' \citep[109]{moody:2003}. Yet, on the other hand, a cohesive group has multiple independent relational paths among all pairs of members. According to the formal definition ``a group's structural cohesion is equal to the minimum number of independent paths linking each pair of actors in the group'' \citep[109]{moody:2003}. These two definitions are mathematically equivalent in terms of the graph theoretic concept of connectivity as defined by Menger's Theorem \citep[330]{white:2001}, which can be formulated locally: ``The minimum node cut set $\kappa(u,v)$ separating a nonadjacent $u,v$ pair of nodes equals the maximum number of node-independent $u-v$ paths''; and globally: ``A graph is $k$-connected if and only if any pair of nodes $u,v$ is joined by at least $k$ node-independent $u-v$ paths''. Thus Menger's theorem links with an equivalence relation a structural property of graphs ---connectivity based on cut sets--- with how graphs are traversed ---the number of node independent paths among pairs of different nodes. This equivalence relation has a deep sociological meaning because it allows for the definition of structural cohesion in terms of the difficulty to pull a group apart by removing actors and, at the same time, in terms of multiple relations between actors that keep a group together.

The starting point of cohesion in a social group is a state where every actor can reach every other actor through at least one relational path. The emergence of a giant component ---a large set of nodes in a network that have at least one path that links any two nodes--- is a minimal condition for the development of group cohesion and social solidarity. \citet{moody:2003} argue that, in this situation, the removal of only one node can affect the flow of knowledge, information and resources in a network because there is only one single path that links some parts of the network. Thus, if a network has actors who are articulation points, their role in keeping the network together is critical; and by extension the network can be disconnected by removing them.  \citet{moody:2003} convincingly argue that biconnectivity provides a baseline threshold for strong structural cohesion in a network because its cohesion does not depend on the presence of any individual actor and the flow of information or resources does not need to pass through a single point to reach any part of the network. Therefore, the concept of robustness is at the core of the structural cohesion approach to subgroup cohesion.

Note that the bicomponent structure of a graph is an exact partition of its edges, which means that each edge belongs to one, and only one, bicomponent; but this is not the case for nodes because $k$-components can overlap in $k-1$ nodes. In the case of bicomponents, articulation points belong to all bicomponents that they separate. Thus, this formalization of subgroup cohesion allows limited horizontal overlapping over $k$-components of the same $k$. On the other hand, the $k$-component structure of a network is inherently hierarchical because $k$-components are nested in terms of connectivity: a connected graph can contain several 2-components, each of which can contain one or more tricomponents, and so forth. This is one of the bases over which the structural cohesion model is built and it is specially useful for operationalizing the hierarchical conception of nested social groups.

However, one shortcoming of classifying cohesive subgroups only in terms of node connectivity is that $k$-components of the same $k$ are always considered equally cohesive despite the fact that one of them might be very close to the next connectivity level, while the other might barely qualify as a component of level $k$ (i.e. removing a few edges could reduce the connectivity level to $k - 1$). \citet{white:2001} propose to complement node connectivity with the measure of conditional density. If a subgroup has node connectivity $k$, then its internal density can only vary within a limited range if the subgroup maintains that same level of connectivity. Thus, they propose to combine node connectivity and conditional density to have a continuous measure of cohesion.  But connectivity is a better measure than density for measuring cohesion because there is no guarantee that a denser subgroup is more robust to node removal than a sparser one, given that both have the same node connectivity $k$.

Building on this insight, we propose using another connectivity-based metric to obtain a continuous and more granular measure of cohesion: the average node connectivity. Node connectivity is a measure based on a worst-case scenario in the sense that to actually break apart a $k$ connected graph by only removing $k$ nodes we have to carefully choose which nodes to remove. Recent work on network robustness and reliability \citep*{albert:2000,dodds:2003} use as the main benchmark for robustness the tolerance to the random or targeted removal of nodes by degree; it is unlikely that by using either of these attack tactics we could disconnect a $k$ connected graph by only removing $k$ nodes. Thus node connectivity does not reflect the typical impact of removing nodes in the global connectivity of a graph $G$. \citet*{beineke:2002} propose the measure of \emph{average node connectivity} of $G$, denoted $\bar{\kappa}(G)$, defined as the sum of local node connectivity between all pairs of different nodes of $G$ divided by the number of distinct pairs of nodes. Or put more formally:

\begin{equation}
\bar{\kappa}(G) = \frac{\sum_{u,v} \kappa_{G}(u,v)}{{n \choose 2}}
\end{equation}

Where $n$ is the number of nodes of $G$. In contrast to node connectivity $\kappa$, which is the minimum number of nodes whose removal disconnects some pairs of nodes, the average connectivity $\bar{\kappa}(G)$ is the expected minimal number of nodes that must be removed in order to disconnect an arbitrary pair of nodes of $G$. For any graph $G$ it holds that $\bar{\kappa}(G) \ge \kappa(G)$. As \citeauthor{beineke:2002} show, average connectivity does not increase only with the increase in the number of edges: graphs with the same number of nodes and edges, and the same degree for each node can have different average connectivity \citep[figure 2, 33]{beineke:2002}. Thus, this continuous measure of cohesion doesn't have the shortcomings of conditional density to measure the robustness of the cohesive subgroups.

The relation between node connectivity and average node connectivity is analog to the relation between diameter and average distance. The diameter of a graph $G$ is the maximum distance between any two nodes of $G$, and like node connectivity, it is a worst-case scenario. It does not reflect the typical distance that separates most pairs of nodes in $G$. When modeling distances between actors in networks, it is better to use the average path length ($L$) because it is close to the typical case: if we choose at random two nodes from a network, it is more likely that their distance is closer to the average than to the maximum distance. Taking into account the average connectivity of each one of the $k$-components of a network allows a more fine grained conception of structural cohesion because, in addition to considering the minimum number of nodes that must be removed in order to disconnect a subgroup, we also consider the number of nodes that, on average, have to be removed to actually disconnect an arbitrary pair of nodes of the subgroup. The latter is a better measure of subgroup robustness than the departure of key individuals from the network.

Structural cohesion is a powerful explanatory factor for a wide variety of interesting empirical social phenomena. It can be used to explain, for instance: the likelihood of building alliances and partnerships among biotech firms \citep{powell:2005}; how positions in the connectivity structure of the Indian inter-organizational ownership network are associated with demographic features (age and industry); and differences in the extent to which firms engage in multiplex and high-value exchanges \citep{mani:2014}. Social cohesion can also help us understand degrees of school attachment and academic performance in young people, as well as the tendency of firms to enroll in similar political activity behaviors \citep{moody:2003}.  It offers insight, also, into emerging trust relations among neighborhood residents or the hiring relations among top level US graduate programs \citep{grannis:2009}. In addition to social solidarity and group cohesion, the model can equally fit many relevant theoretical issues, such as conceptualizing structural differences among fields and organizations \citep{white:2004}, operationalizing the structural component of social embeddedness \citep{granovetter:1985,moody:2004}, explaining the role of highly connected subgroups in boosting diffusion in social networks without a high rate of decay \citep{moody:2004, white:2001}, or highlighting the complexity and diversity of the structure of real world markets beyond stylized one-dimensional characterizations of the market \citep{mani:2014}.

Despite all its merits, the structural cohesion model has not been widely applied to empirical analysis because it is not practical to compute it for networks with more than a few thousands nodes and edges due to its computational complexity. What's more, it is not implemented in most popular network analysis software packages. In the next section, we will review the existing algorithm to compute the $k$-component structure for a given network, before introducing our heuristics to speed up the computation.

\section{Existing algorithms for computing $k$-component structure}

\citet[appendix A]{moody:2003} provide an algorithm for identifying $k$-components in a network, which is based on the \citet{kanevsky:1993} algorithm for finding all minimum-size node cut-sets of a graph; i.e. the set (or sets) of nodes of cardinality $k$ that, if removed, would break the network into more connected components. The algorithm consists of 4 steps:

\begin{enumerate}

\item Identify the node connectivity, $k$, of the input graph using flow-based connectivity algorithms \citep[chapter 7]{brandes:2005}.

\item Identify all $k$-cutsets at the current level of connectivity using the \citet{kanevsky:1993} algorithm.

\item Generate new graph components based on the removal of these cutsets (nodes in the cutset belong to both sides of the induced cut).

\item If the graph is neither complete nor trivial, return to 1; otherwise end.

\end{enumerate}

As the authors note, one of the main strengths of the structural cohesion approach is that it is theoretically applicable to both small and large groups, which contrasts with the historical focus of the literature on small groups when dealing with cohesion. But the fact that this concept and the algorithm proposed by the authors, are theoretically applicable to large groups does not mean that this would be a practical approach for analyzing the structural cohesion on large social networks \footnote{The fastest implementation of this algorithm runs in $O(N^4)$ time \citep{igraph} which is impractical for moderately large networks.}.

The equivalence relation established by Menger's theorem between node cut sets and node independent paths can be useful to compute connectivity in practical cases but both measures are almost equally hard to compute if we want an exact solution. However, \citet{white:2001b} proposed a fast approximation algorithm for finding good lower bounds of the number of node independent paths between two nodes. This smart algorithm is based on the idea of searching paths between two nodes, marking the nodes of the path as ``used'' and searching for more paths that do not include nodes already marked. But instead of trying all possible paths without order, this algorithm considers only the shortest paths: it finds node independent paths between two nodes by computing their shortest path, marking the nodes of the path found as ``used'' and then searching other shortest paths excluding the nodes marked as ``used'' until no more paths exist. Because finding the shortest paths is faster than finding other kinds of paths, this algorithm runs quite fast, but is not exact because a shortest path could use nodes that, if the path were longer, may belong to two different node independent paths \citep[section III]{white:2001b}. Therefore a condition for the use of this approximation algorithm would be that the networks analyzed should be sparse; this will reduce its inaccuracy because it will be less likely that a shorter path uses nodes that could belong to two or more longer node independent paths.

\citeauthor{white:2001b} suggest that this algorithm could be used to find $k$-components.  First one should compute the node independent paths between all pairs of different nodes of the graph. Then build an auxiliary graph in which two nodes are linked if they have at least $k$ node independent paths connecting them. The induced subgraph of all nodes of each connected component of the auxiliary graph form an extra-cohesive block of level $k$ (like a $k$-component but with the difference that not all node independent paths run entirely inside the subgraph). Finally, we could approximate the $k$-component structure of a graph by successive iterations of this procedure.

However, there are a few problems with this approach. First, a $k$-component is defined as a maximal subgraph in which all pairs of different nodes have, at least, $k$ node independent paths between them. If we rely on the connected components of the auxiliary graph as proposed by \citet{white:2001b} we will include in a given $k$-component all nodes that have at least $k$ node independent paths with \emph{only one} other node of the subgraph. Thus, the cohesive subgraphs detected won't have to be $k$-components as defined in graph theory. Second, $k$-components can overlap in $k-1$ nodes. If we only consider connected components (i.e. $1$-components) in the auxiliary graph, we will not be able to distinguish overlapping $k$-components. Finally, the approach proposed by \citeauthor{white:2001b} is not practical in computational terms for large networks because of its recursive nature and because it needs to compute node independent paths for all pairs of different nodes in the network as starting point.

\section{Heuristics for computing $k$-components and their average connectivity}

The logic of the algorithm presented here is based on repeatedly applying fast algorithms for $k$-cores \citep{batagelj:2011} and biconnected components \citep{tarjan:1972} in order to narrow down the number of pairs of different nodes over which we have to compute their local node connectivity for building the auxiliary graph in which two nodes are linked if they have at least $k$ node independent paths connecting them. We follow the classical insight that, ``$k$-cores can be regarded as seedbeds, within which we can expect highly cohesive subsets to be found'' \citet[281]{seidman:1983}. More formally, our approach is based on Whitney's theorem \citep[328]{white:2001}, which states an inclusion relation among node connectivity $\kappa(G)$, edge connectivity $\lambda(G)$ and minimum degree $\delta(G)$ for any graph $G$:

\begin{equation}
\kappa(G) \le \lambda(G) \le \delta(G)
\end{equation}

This theorem implies that every $k$-component is nested inside a $k$-edge-component, which in turn, is contained in a $k$-core. This approach, unlike the proposal of \citet{white:2001b}, does not require computing node independent paths for all pairs of different nodes as a starting point, thus saving an important amount of computation. Moreover it does not require recursively applying the same procedure over each subgraph. In our approach we only have to compute node independent paths among pairs of different nodes in each biconnected part of each $k$-core, and repeat this procedure for each $k$ from 3 to the maximal core number of a node in the input network.

The aim of the heuristics presented here is to provide a fast and reasonably accurate way of analyzing the cohesive structure of empirical networks of thousands of nodes and edges. As we have seen, $k$-components are the cornerstone of structural cohesion analysis. But they are very expensive to compute. Our approach consists of computing extra-cohesive blocks of level $k$ for each biconnected component of a $k$-core. Extra-cohesive blocks are a relaxation of the $k$-component concept in which not all node independent paths among pairs of different nodes have to run entirely inside the subgraph. Thus, there is no guarantee that an extra-cohesive block of level $k$ actually has node connectivity $k$. We introduce an additional constraint to the extra-cohesive block concept in order to approximate $k$-components: our algorithm computes extra-cohesive blocks of level $k$ that are also $k$-cores by themselves in $G$. Based on several tests with synthetic and empirical networks presented below, we show that usually extra-cohesive blocks detected by our algorithm have indeed node connectivity $k$. Futhermore, extra-cohesive blocks maintain high requirements in terms of multiconnectivity and robustness, thus conserving the most interesting properties from a sociological perspective on the structure of social groups.

Combining this logic with three observations about the auxiliary graph $H$ allows us to design a new algorithm for finding extra-cohesive blocks in each biconnected component of a $k$-core, that can either be exact but slow ---using flow-based algorithms for local node connectivity \citep[Chapter 7]{brandes:2005}--- or fast and approximate, giving a lower bound with certificate of the composition and the connectivity of extra-cohesive blocks ---using \citet{white:2001b} approximation for local node connectivity. Once we have a fast way to compute extra-cohesive blocks, we can approximate $k$-components by imposing that the induced subgraph of the nodes that form an extra-cohesive block of $G$ have to also be a $k$-core in $G$.

Let $H$ be the auxiliary graph in which two nodes are linked if they have at least $k$ node independent paths connecting them in each of the biconnected components of the core of level $k$ of original graph $G$ (for $k > 2$). The first observation is that complete subgraphs in $H$ ($H_{clique}$) have a one to one correspondence with subgraphs of $G$ in which each node is connected to every other node in the subgraph for at least $k$ node independent paths. Thus, we have to search for cliques in $H$ in order to discover extra-cohesive blocks in $G$.

The second observation is that an $H_{clique}$ of order $n$ is also a core of level $n-1$ (all nodes have core number $n-1$), and the degree of all nodes is also $n-1$. The auxiliary graph $H$ is usually very dense, because we build a different $H$ for each biconnected part of the core subgraph of level $k$ of the input graph $G$. In this kind of network big clusters of almost fully connected nodes are very common. Thus, in order to search for cliques in $H$ we can do the following:

\begin{enumerate}

\item For each core number value $c_{value}$ in each biconnected component of $H$:

\item Build a subgraph $H_{candidate}$ of $H$ induced by the nodes that have \emph{exactly} core number $c_{value}$. Note that this is different than building a $k$-core, which is a subgraph induced by all nodes with core number \emph{greater or equal than} $c_{value}$.

\item If $H_{candidate}$ has order $c_{value} + 1$ then it is a clique and all nodes will have degree $n - 1$. Return the clique and continue with the following candidate.

\item If this is not the case, then some nodes will have degree $< n - 1$. Remove all nodes with minimum degree from $H_{candidate}$.

\item If the graph is trivial or empty, continue with the following candidate. Or otherwise recompute the core number for each node and go to 3.

\end{enumerate}

Finally, the third observation is that if two $k$-components of different order overlap, the nodes that overlap belong to both cliques in $H$ and will have core numbers equal to all other nodes in the bigger clique. Thus, we can account for possible overlap when building subgraphs $H_{candidate}$ (induced by the nodes that have \emph{exactly} core number $c_{value}$) by also adding to the candidate subgraph the nodes in $H$ that are connected to all nodes that have \emph{exactly} core number $c_{value}$. Also, if we sort the subgraphs $H_{candidate}$ in reverse order (starting from the biggest), we can skip checking for possible overlap for the biggest.

Based on these three observations, our heuristics for approximating the cohesive structure of a network and the average connectivity of each individual block, consists of: 

Let $G$ be the input graph. Compute the core number of each node in $G$. For each $k$ from 3 to the maximum core number build a $k$-core subgraph $G_{k-core}$ with all nodes in $G$ with core level $\ge k$.

For each biconnected component of $G_{k-core}$:

\begin{enumerate}

\item Compute local node connectivity $\kappa(u,v)$ between all pairs of different nodes. Optionally store the result for each pair. Either use a flow-based algorithm (exact but slow) or White and Newman's approximation for local node connectivity (approximate but a lot faster).

\item Build an auxiliary graph $H$ with all nodes in this bicomponent of $G_{k-core}$ with edges between two nodes if $\kappa(u,v) \ge k$. For each biconnected component of $H$:

\item Compute the core number of each node in $H_{bicomponent}$, sort the values in reverse order (biggest first), and for each value $c_{value}$:

\begin{enumerate}

\item Build a subgraph $H_{candidate}$ induced by nodes with core number \emph{exactly} equal to $c_{value}$ plus nodes in $H$ that are conected with all nodes with core number equal to $c_{value}$.

\begin{enumerate}
\item If $H_{candidate}$ has order $c_{value} + 1$ then it is a clique and all nodes will have degree $n - 1$. Build a core subgraph $G_{candidate}$ of level $k$ of $G$ induced by all nodes in $H_{candidate}$ that have core number $\ge k$ in G.

\item If this is not the case, then some nodes will have degree $< n - 1$. Remove all nodes with minimum degree from $H_{candidate}$. Build a core subgraph $G_{candidate}$ of level $k$ of $G$ induced by the remaining nodes of $H_{candidate}$ that have core number $\ge k$ in G. 

\begin{enumerate}

\item If the resultant graph is trivial or empty, continue with the following candidate.

\item Else recompute the core number for each node in the new $H_{candidate}$ and go to (i).

\end{enumerate}

\end{enumerate}

\item The nodes of each biconnected component of $G_{candidate}$ are assumed to be a $k$-component of the input graph if the number of nodes is greater than $k$.

\item Compute the average connectivity of each detected $k$-component. Either use the value of $\kappa(u,v)$ computed in step 1 or recalcualte $\kappa(u,v)$ in the induced subgraph of candidate nodes.

\end{enumerate}

\end{enumerate}

Notice that because our approach is based on computing node independent paths between pairs of different nodes, we are able to use these computations to calculate both the cohesive structure and the average node connectivity of each detected $k$-component. Of course, computing average connectivity comes with a cost: either more space to store $\kappa(u,v)$ in step 1, or more computation time in step 3.c if we did not store $\kappa(u,v)$. This is not possible when applying the exact algorithm for $k$-components proposed by \citet{moody:2003} because it is based on repeatedly finding $k$-cutsets and removing them, thus it does not consider node independent paths at all.

The output of these heuristics is an approximation to $k$-components based on extra-cohesive blocks. We find extra-cohesive blocks and not $k$-components because we only build the auxiliary graph $H$ one time on each bicoennected component of a core subgraph of level $k$ from the input graph $G$. Local node connectivity is computed in a subgraph that might be larger than the final $G_{candidate}$ and thus some node independent paths that shouldn't could end up being counted. 

Accuracy can be improved by rebuilding $H$ from the pairwise node connectivity in $G_{candidate}$ and following the remaining steps of the heuristics at the cost of slowing down the computation. There is a trade-off between speed and accuracy. After some tests we decided to compute $H$ only once and lean towards the speed pole of the trade-off. Our goal is to have an usable procedure for analyzing networks of thousands of nodes and edges in which we have substantive interests. Following this goal, the use of \citet{white:2001b} approximation algorithm for local node connectivity in step 3.b is key. It is almost on order of magnitude faster than the exact flow-based algorithms. As usual, speed comes with a cost in accuracy: \citet{white:2001b} algorithm provides a strict lower bound for the local node connectivity. Thus, by using it we can miss an edge in $H$ that should be there. Therefore, a node belonging to a $k$-component could be excluded by the algorithm if we use \citet{white:2001b} approximation in step 3.b . This is a source of false negatives in the process of approximating the $k$-component structure of a network. However, as we discussed above, the inaccuracy of this algorithm for sparse networks in reduced because in those networks the probability that a short node independent path uses nodes that could belong to two or more longer node independent paths is low.

Our tests reveal that the use of \citet{white:2001b} approximation does indeed underestimate the order of some $k$-components, particularly in not very sparse networks. One approach to mitigate this problem is to relax the strict cohesion requirement of $H_{candidate}$ being a clique. Following the network literature on cliques, we can relax its cohesion requirements in terms of degree, coreness and density. We did some experiments and found that a good relaxation criteria is to set a density threshold of 0.95 for $H_{candidate}$; it doesn't increase false positives and does decrease the false negatives derived from the underestimation of local node connectivity of \citet{white:2001b} algorithm.  Other possible criteria that has given good results in our tests is permitting a variation in degree of 2 in $H_{candidate}$ ---that is, that the absolute difference of the maximum an the minimum degree in $H_{candidate}$ is at most 2. The former relaxation criteria is used for all analysis presented below and in the appendix.

This algorithm can be easily generalized so as to be applicable to directed networks provided that the implementation of White and Newman's approximation for pairwise node independent paths supports directed paths (which is the case in our implementation of this algorithm on top of NetworkX library). The only change needed then is to use strongly connected components instead of bicomponents. And, in step 3, to start with core number 2 instead of 3.

In appendix \ref{illustration} we present an illustration of the heuristics using a convenient small synthetic network. In appendix \ref{performance} we present an analysis of the performance of the heuristics compared to the performance of the exact algorithm for finding $k$-components \citep{moody:2003}. In appendix \ref{implementation} we discuss the implementation details of the heuristics; and in appendix \ref{code} we present the python code of our implementation of the heuristics for illustrative purposes\footnote{The fully functional Python code is available from the authors}.

\section{Structural cohesion in collaboration networks}

The structural cohesion model can be used to explain cooperation in different kinds of collaboration networks; for instance, coauthorship networks \citep{moody:2004, white:2004} and collaboration among biotech firms \citep{powell:2005}. Most collaboration networks are bipartite because the collaboration of individuals has as a result ---or, at least, as a relevant byproduct--- some kind of object or event to which its authors are related. All these papers follow the usual practice to deal with two-mode networks: focus the analysis only on one-mode projections. As such, we don't know how much information about their cohesive structure we lose by ignoring the underlying bipartite networks. Recent literature on two-mode networks strongly suggests that it is necessary to analyze two-mode networks directly to get an accurate picture of their structure. For instance, in small world networks, we do know that focusing only on projections overestimates the smallworldiness of the network \citep{uzzi:2007}.  We also know that generalizing clustering coefficients to bipartite networks can offer key information that is lost in the projection  \citep{robins:2004,lind:2005,opsahl:2011}.  Finally,  the loss of information is also critical in many other common network measures: degree distributions, density, and assortativity \citep{latapy:2008}. We show that this is also the case for the $k$-component structure of collaboration networks.

Structural cohesion analysis based on the $k$-component structure of bipartite networks has been conducted very rarely and only on very small networks \citep{white:2004}. The limited diffusion of these studies can be readily explained by the fact that bipartite networks are usually quite a lot bigger than their one-mode counterparts, and the computational requirements, once again, stifled empirical research in this direction. Other measures have been developed to deal with cohesion in large bipartite networks, such as $(p, q)$-cores or 4-ring islands \citep{ahmed:2007}. However, the former is a bipartite version of $k$-cores and thus it has the same limitations for subgroup identification; while the latter is very useful to determine subgraphs in large networks that are more strongly connected internally than with the rest of the network, but also lacks some of the key elements of the definition for groups in the sociological literature, such as being hierarchical and allowing for overlaps.

\begin{table}[h]
\begin{center}
\begin{small}
\begin{tabular}{|c|c|c|c|c|c|c|c|c|}
\hline
&\multicolumn{4}{|c|}{Bipartite}&\multicolumn{4}{|c|}{Unipartite}\\
Network&\# nodes&\# edges&Av. degree&Time(s)&\# nodes&\# edges&Av. degree&Time(s)\\
\hline
%&&&&&&&&\\
%Debian Etch&11583&17522&3.03&522.0&1317&7528&11.43&148.4\\
Debian Lenny&13121&20220&3.08&1105.2&1383&5216&7.54&204.7\\
High Energy (theory)&26590&37566&2.81&3105.7&9767&19331&3.97&7136.0\\
Nuclear Theory&10371&15969&3.08&1205.2&4827&14488&6.00&3934.1\\
\hline
\end{tabular}
\end{small}
\caption{Collaboration networks analyzed from science and from software development. See text for details on their content. Time refers to the execution of our heuristics on each network expressed in seconds.}
\label{desc}
\end{center}
\end{table}

The heuristics for structural cohesion presented here allows us to compute connectivity-based measures on large networks (up to tens of thousands of nodes and edges) quickly enough to be able to build suitable null models.  Furthermore we will be able to compare the results for bipartite networks with their one-mode projections. To illustrate those points we use data on collaboration among software developers in one organization (the Debian project) and scientists publishing papers in the arXiv.org electronic repository in two different scientific fields: High Energy Theory and Nuclear Theory. We built the Debian collaboration network by linking each software developer with the packages (i.e. programs) that she uploaded to the package repository of the Debian Operating System during a complete release cycle. We analyze the Debian Operating System version 5.0, codenamed ``Lenny'', which was developed from April 8, 2007, to February 1, 2009. Scientific networks are built using all the papers uploaded to the arXiv.org preprint repository from January 1, 2006, to December 31, 2010, for two well established scientific fields: High Energy Physics Theory and Nuclear Theory. In these networks each author is linked to the papers that she has authored during the time period analyzed. One-mode projections are always on the human side: scientists linked together if they have coauthored a paper, and developers linked together if they have worked on the same program. Table \ref{desc} presents some details on those networks.

In the remaining part of this section we perform three kinds of analysis to demonstrate the loss of information we incur when focusing only on one-mode projections when dealing with bipartite networks. First, we present a tree representation of the $k$-component structure ---the cohesive blocks structure \citep{white:2001,moody:2003,white:2004,mani:2014}--- for our bipartite networks and their one-mode projections, both for actual networks and for their random counterparts. Second, we present a comparison among actual and random networks (both for one and two-mode) on the $k$-number frequencies of nodes. Finally, we present a novel graphic representation of the structural cohesion of a network, based on three-dimensional scatter plot, using average node connectivity as a synthetic and more informative measure of cohesion of each $k$-component.

For the first two analyses we do need to generate null models in order to discount the possibility that the observed structure of actual networks is just the result of randomly mixing papers and scientists or packages and developers. The null models used in this paper are based on a bipartite configuration model \citep{newman:2003}, which consists of generating networks by randomly assigning papers/programs to scientists/developers but maintaining constant the distribution of papers per scientists and scientists by paper observed in the actual networks, that is the bipartite degree distribution. For one-mode projections, we generated bipartite random networks based on their original bipartite degree distribution, and then performed the one-mode projection. This is a common technique for avoiding overestimating the local clustering of one-mode projections \citep{uzzi:2007}. As the configuration model can generate some multiple edges and self-loops, we followed the usual practice of deleting them before the analysis in order to guarantee that random networks are simple, like actual networks.

So let's start with the tree representation of the cohesive blocks structure. As proposed by \citet{white:2004}, we can represent the $k$-component structure of a network by drawing a tree whose nodes are $k$-components; two nodes are linked if the $k$-component of higher level is nested inside the $k$-component of lower level (see \citet[1643,1651]{mani:2014} for this kind of analysis on the Indian interorganizational ownership network). This representation of the connectivity structure can be built during the run time of the exact algorithm. However, because our heuristics are based on finding node independent paths, we have to compute first the $k$-components hierarchy, and then construct the tree that represents the connectivity structure of the network.

\begin{landscape}
\begin{figure}[p]
\centering
\subfloat[Actual 2 mode]{
\label{fig:cb_nucl_2m}
\includegraphics[scale=0.15]{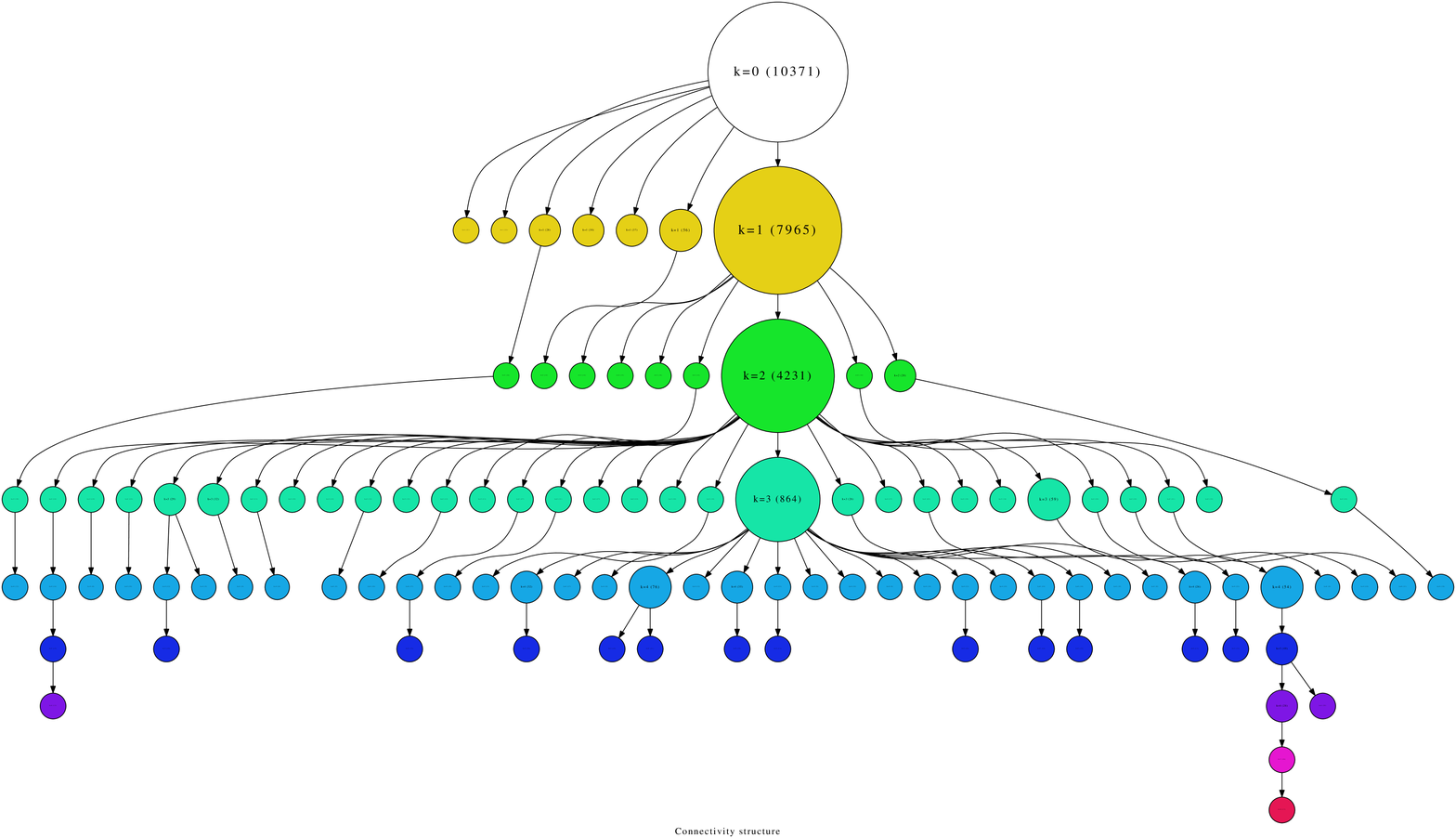}
}
\hspace{.2in}
\subfloat[Random 2 mode]{
\label{fig:cb_nucl_r_2m}
\includegraphics[scale=0.15]{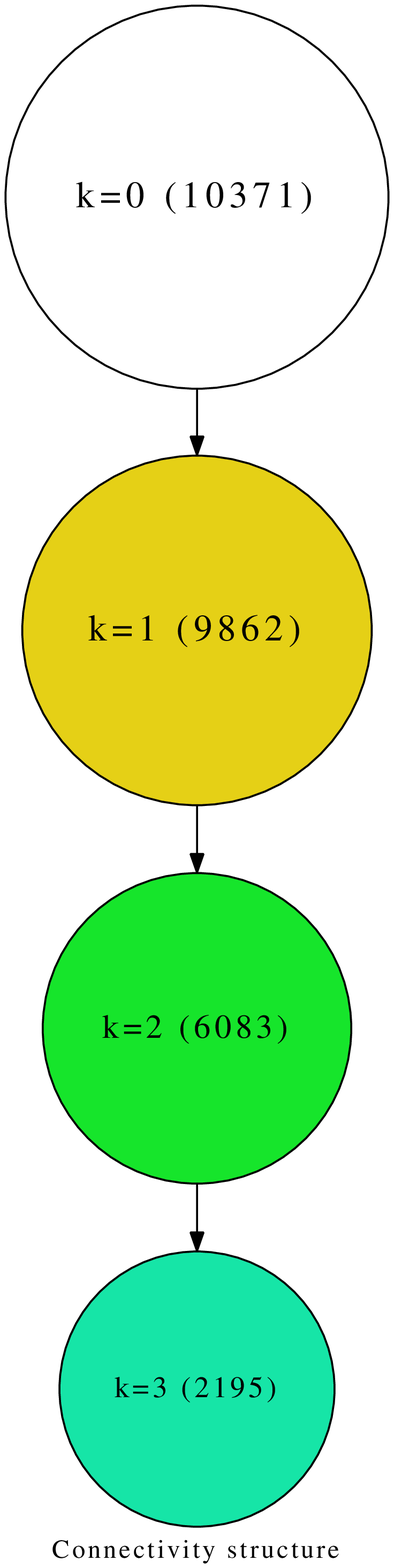}
} 
\hspace{.2in}
\subfloat[Actual 1 mode]{
\label{fig:cb_nucl_1m}
\includegraphics[scale=0.07]{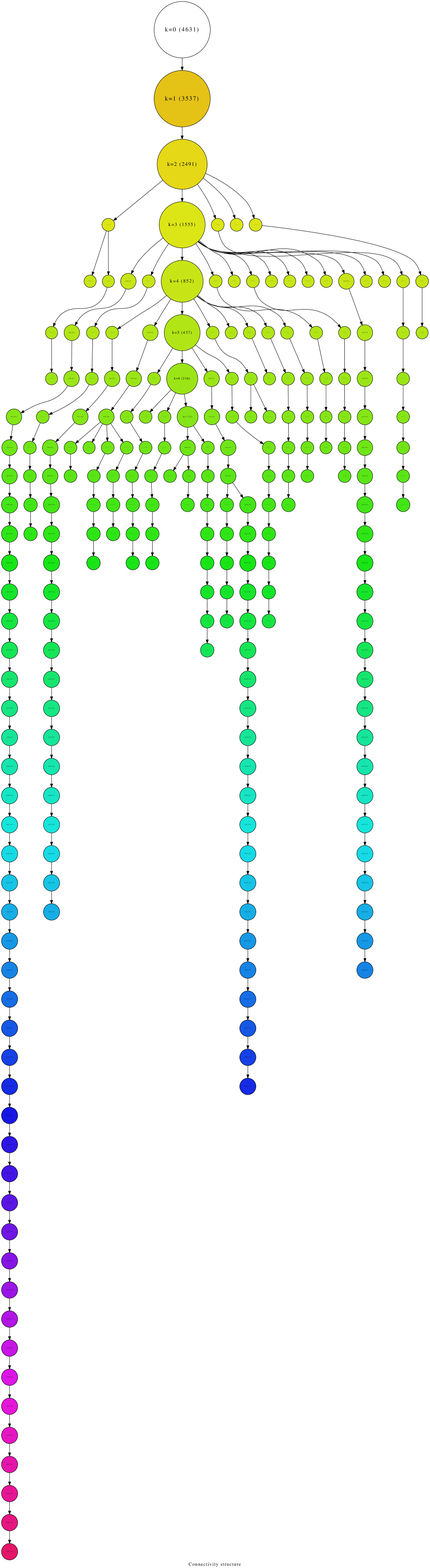}
}
\hspace{.2in}
\subfloat[Random 1 mode]{
\label{fig:cb_nucl_r_1m}
\includegraphics[scale=0.07]{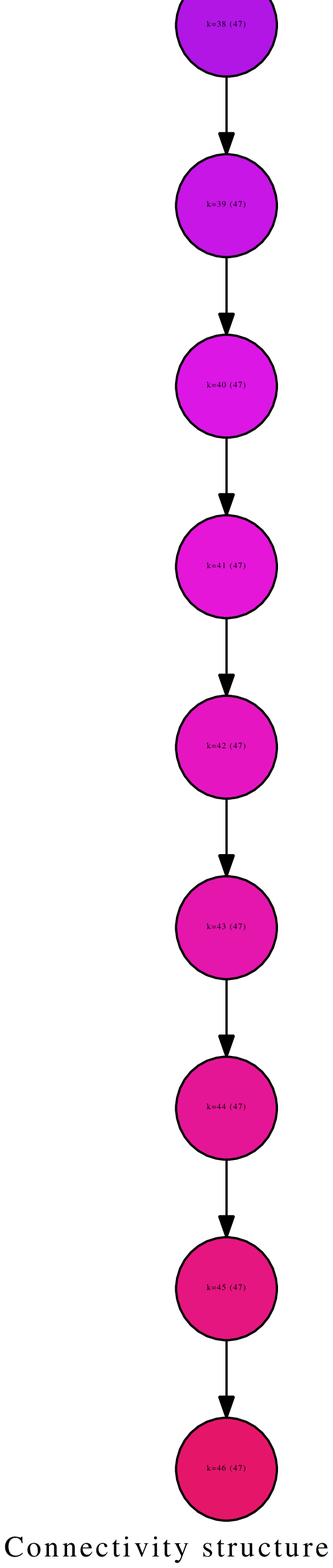}
}

\caption{Cohesive blocks for two-mode and one-mode Nuclear Theory collaboration networks, and for their random counterparts. Random networks were generated using a bipartite configuration model. We built 1000 random networks and chose one randomly, see text for details. For lower connectivity levels we have removed some small $k$-components to improve the readability: we do not show 1-components with less than 20 nodes, 2-components with less than 15 nodes, or tricomponents with less than 10 nodes.}
\label{fig:cb_nucl_all}
\end{figure}
\end{landscape}

% Debian
\begin{landscape}
\begin{figure}[p]
\centering
\subfloat[2 mode]{
\label{fig:cb_lenny_2m}
\includegraphics[scale=0.25]{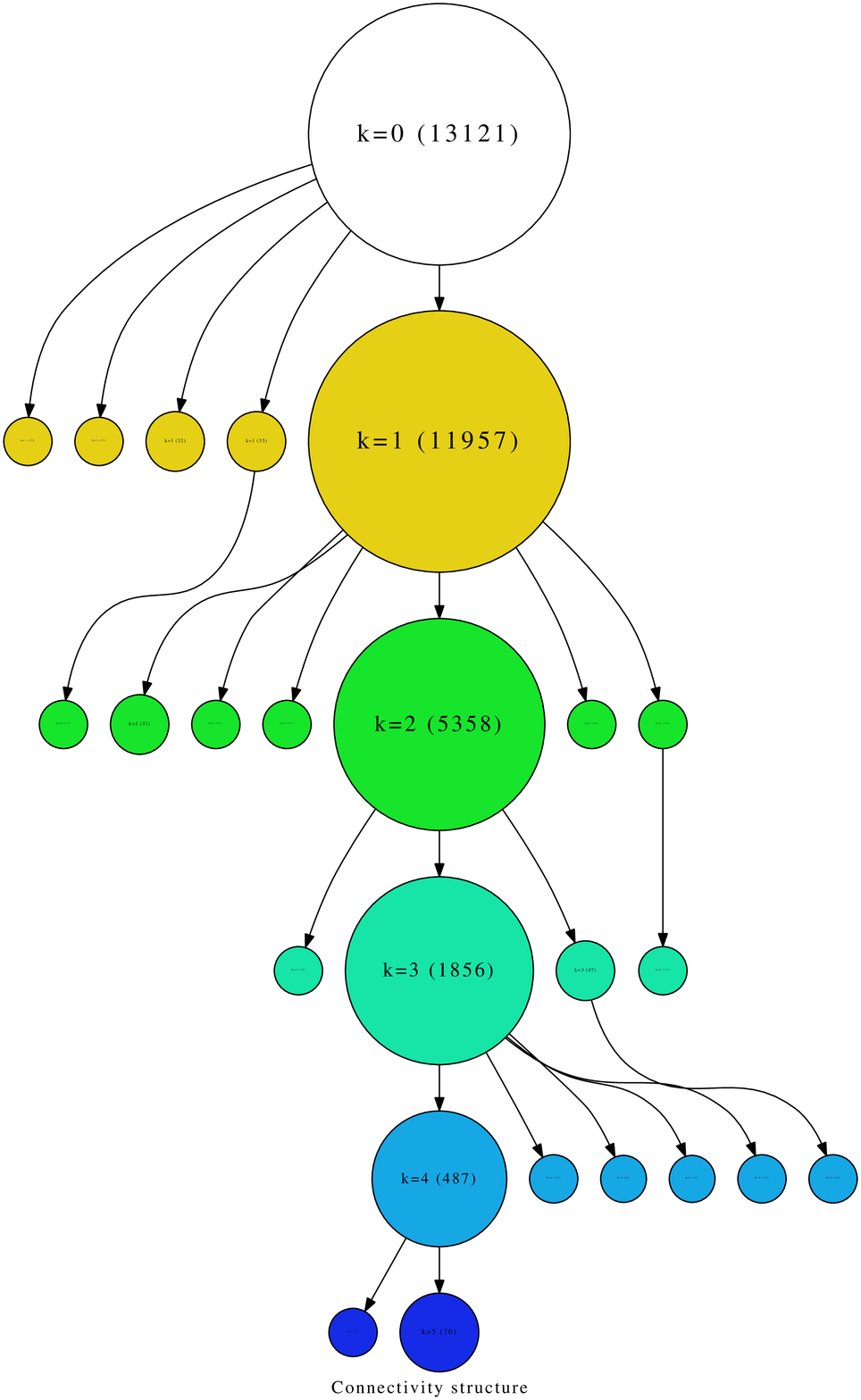}
}
\hspace{.95in}
\subfloat[Random 2 mode]{
\label{fig:cb_lenny_2m_r}
\includegraphics[scale=0.25]{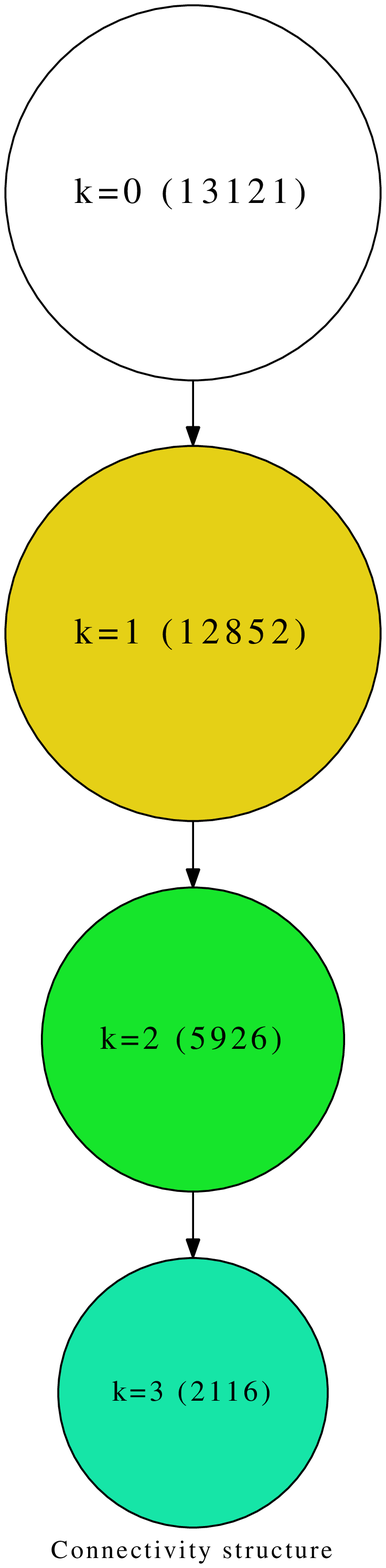}
}
\hspace{.95in}
\subfloat[1 mode]{
\label{fig:cb_lenny_1m}
\includegraphics[scale=0.08]{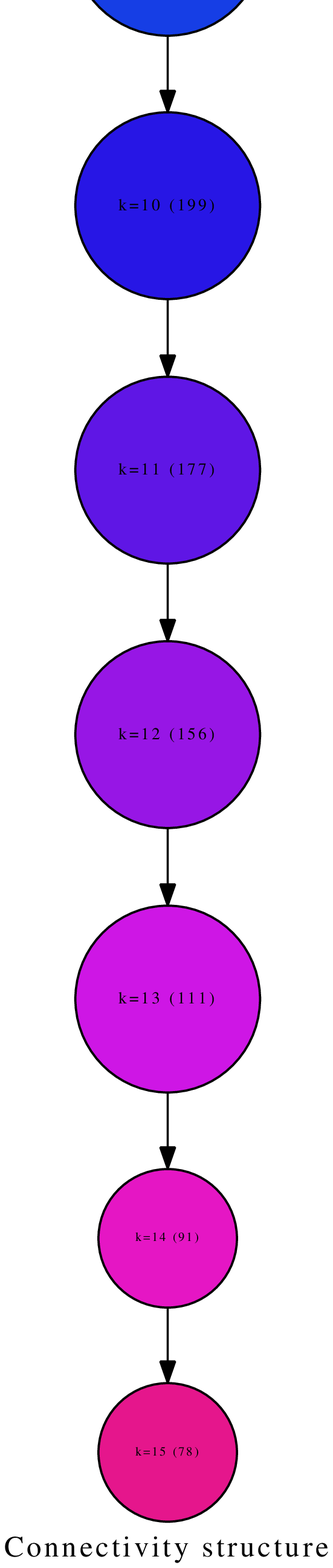}
}
\hspace{.95in}
\subfloat[Random 1 mode]{
\label{fig:cb_lenny_1m_r}
\includegraphics[scale=0.08]{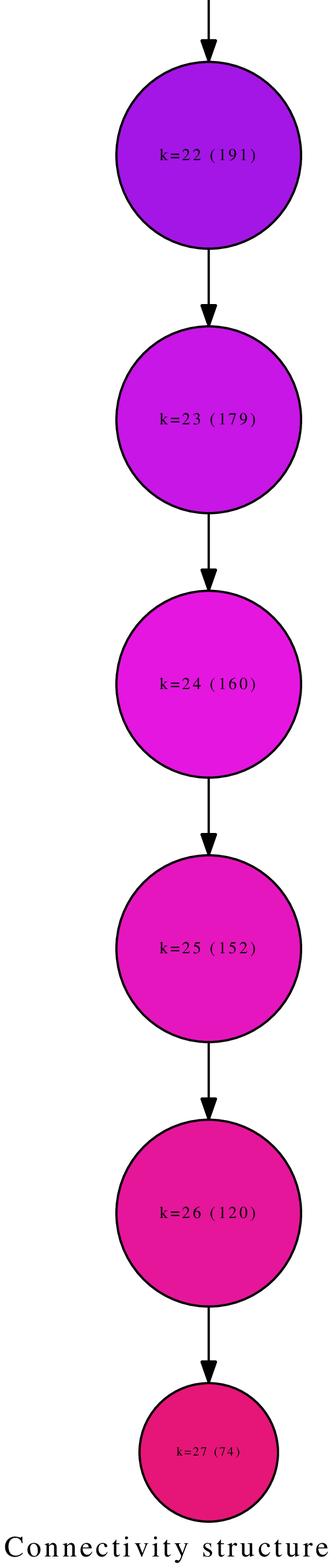}
}
\caption{Cohesive blocks for two-mode and one-mode Debian collaboration networks, and for their random counterparts. Random networks were generated using a bipartite configuration model. We built 1000 random networks and chose one randomly, see text for details. For lower connectivity levels we have removed some small $k$-components to improve the readability: we do not show 1-components with less than 20 nodes, 2-components with less than 15 nodes, or tricomponents with fewer than 10 nodes.}
\label{fig:cb_lenny_all}
\end{figure}
\end{landscape}

Figures \ref{fig:cb_nucl_2m} and \ref{fig:cb_nucl_1m} show the connectivity structure of Nuclear Theory collaboration networks represented as a tree, the former for the two-mode network and the latter for one-mode ones. As we can see, both networks display non-trivial structure. The two-mode network has up to an 8-component, but most nodes are in $k$-components with $k < 6$. Up to $k = 3$ most nodes are in giant $k$-components, but for $k = \{4,5\}$ there are many $k$-components of similar order. Figure \ref{fig:cb_nucl_1m}, which corresponds to the one-mode projection, has a lot more connectivity levels ---a byproduct of the mathematical transformation from two-mode to one-mode. In this network, the maximum connectivity level is 46; the four long legs of the plot correspond to 4 cliques with 47, 31, 27 and 25 nodes. Notice that each one of these 4 cliques are already a separated $k$-component at $k=7$. It is at this level of connectivity ($k=\{7,8\}$) where the giant $k$-components start to dissolve and many smaller $k$-components emerge.

In order to be able to assess the significance of the results obtained, we have to compare the connectivity structure of actual networks with the connectivity structure of a random network that maintains the observed bipartite degree distribution. In this case, we compare actual networks with only one random network. We obtained it by generating 1000 random networks and choosing one randomly. Figures \ref{fig:cb_nucl_r_2m} and \ref{fig:cb_nucl_r_1m} show the connectivity structure of the random counterparts for Nuclear Theory collaboration networks. For the two-mode network, instead of the differentiated connectivity structure displayed by the actual bipartite network, there is a flatter connectivity structure, where the higher level $k$-component is a tricomponent. Moreover, instead of many small $k$-components at high connectivity levels, the random bipartite network has only giant $k$-components where all nodes with component number $k$ are. In this case, the one-mode network is also quite different from its random counterpart. There are only giant $k$-components up until $k=15$, where the four cliques observed in the actual network separate from each other to form distinct $k$-components.

The hierarchy of the connectivity structure displayed in these plots allows us to do meaningful comparisons between networks in terms of their connectivity structure. For instance, figures \ref{fig:cb_lenny_2m} and \ref{fig:cb_lenny_1m} show the connectivity structure of Debian collaboration networks. The former displays the bipartite connectivity structure, which is quite different from two-mode Nuclear Theory structure discussed above. Although there are some small $k$-components for each connectivity level, most of the nodes with $k$-number $k$ are in a giant $k$-component that encompasses most of the nodes of that level. Even at the top level of connectivity ($k=5$), 80 percent of the 88 nodes with $k$-number 5 are in the same 5-component. Figure \ref{fig:cb_lenny_1m} displays the cohesive block structure for its one-mode projection. It consists of a monotonous linear succession of increasingly smaller $k$-components nested inside each other. 

Figures \ref{fig:cb_lenny_2m_r} and \ref{fig:cb_lenny_1m_r} show the connectivity structure of the random counterparts of Debian collaboration networks. The random one-mode projection has the same structure than its actual counterpart, a single long chain of $k$-components nested inside each other. However, the random two-mode structure is quite different from its actual counterpart: it consists of a chain of single cohesive blocks. At lower connectivity levels, up to $k=3$, the random network have more nodes in those giant $k$-components than its actual counterpart; but the actual Debian two-mode network has a bigger 4-component and also 2 5-components that are not present in its random counterpart. Thus, in terms of their connectivity structure, two-mode networks are farther apart from their random counterparts than their one-mode projections.

\begin{figure}[p]
\centering
%\subfloat[Bipartite network formed by developers and packages of 2 years of collaboration (from 2005 to 2007) on the release codenamed Etch of the Debian operating system]{
%\label{fig:etch2m}
%\includegraphics[scale=0.25]{barplot_etch_2mode}
%}
%\hspace{.05in}
%\subfloat[Unipartite network formed by developers of 2 years of collaboration (from 2005 to 2007) on the release codenamed Etch of the Debian operating system]{
%\label{fig:etch1m}
%\includegraphics[scale=0.25]{barplot_etch_1mode}
%}
%\hspace{.01in}
\subfloat[Bipartite network formed by developers and packages during 2 years of collaboration (from 2007 to 2009) on the release codenamed Lenny of the Debian operating system]{
\label{fig:lenny2m}
\includegraphics[scale=0.35]{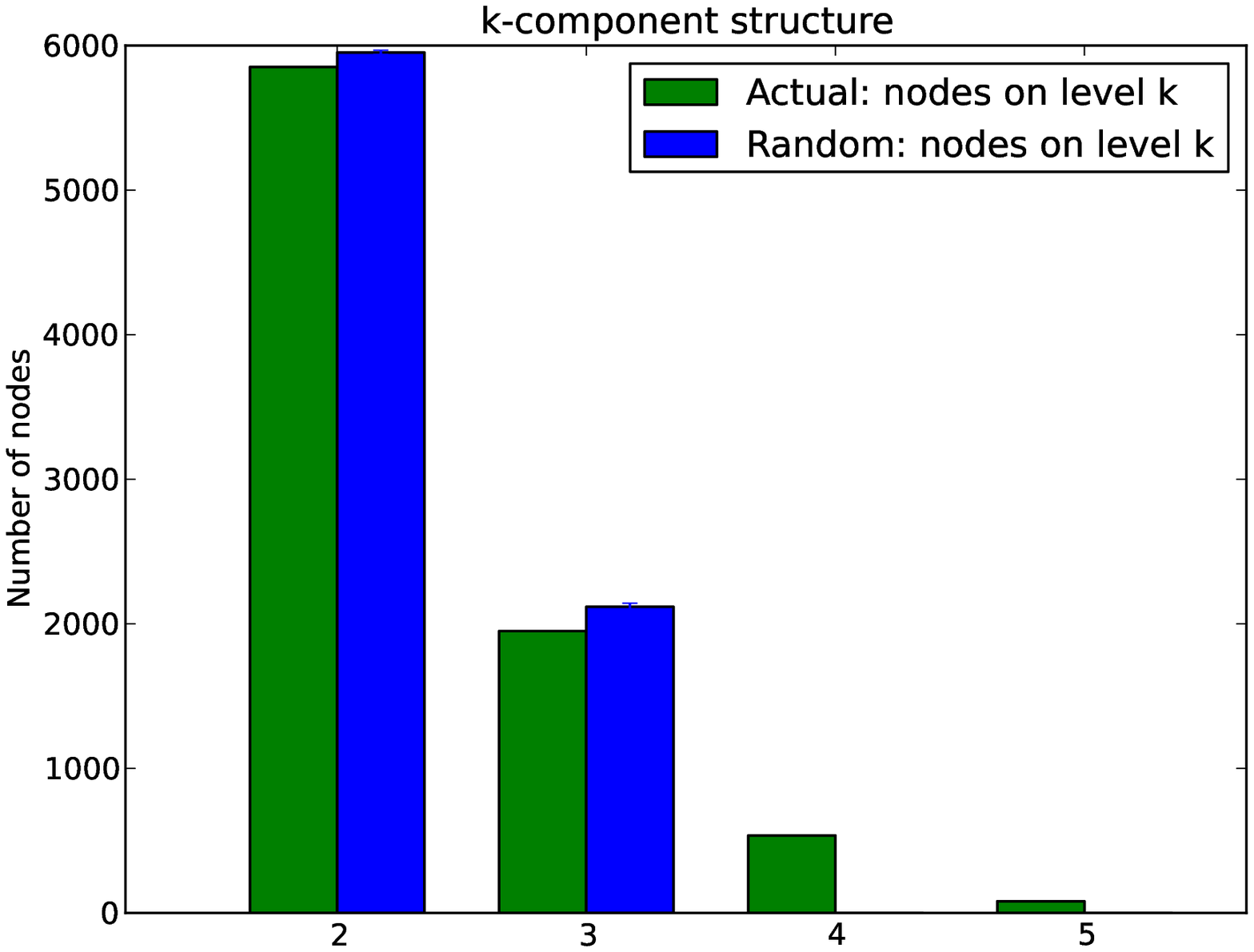}
}
\hspace{.05in}
\subfloat[Unipartite network formed by developers during 2 years of collaboration (from 2007 to 2009) on the release codenamed Lenny of the Debian operating system]{
\label{fig:lenny1m}
\includegraphics[scale=0.27]{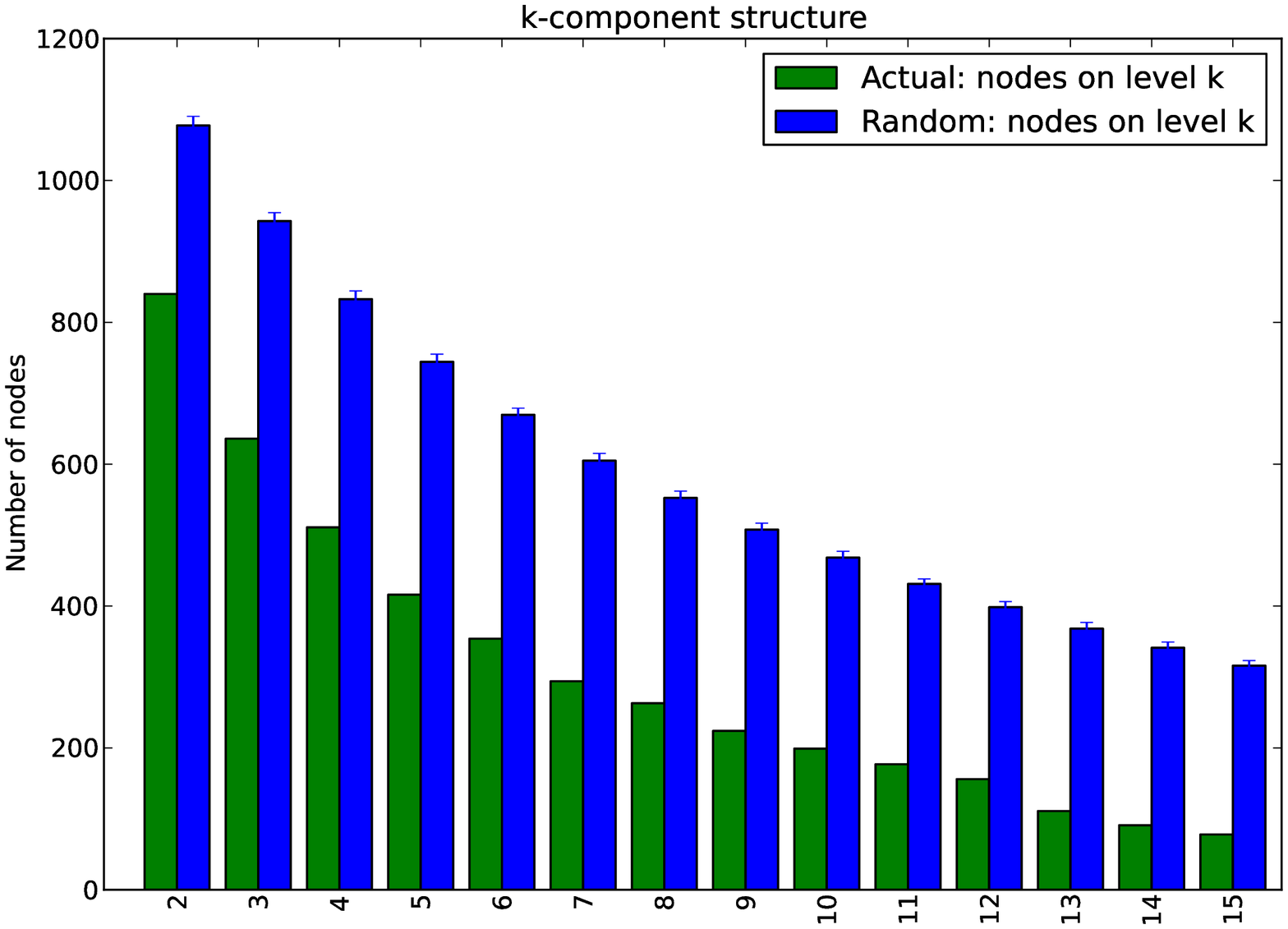}
}
\hspace{.01in}
\subfloat[Bipartite network formed by scientists and preprints during 5 years (2006-2010) in the high energy physics (theory) section of arXiv.org]{
\label{fig:hep_th_2m}
\includegraphics[scale=0.35]{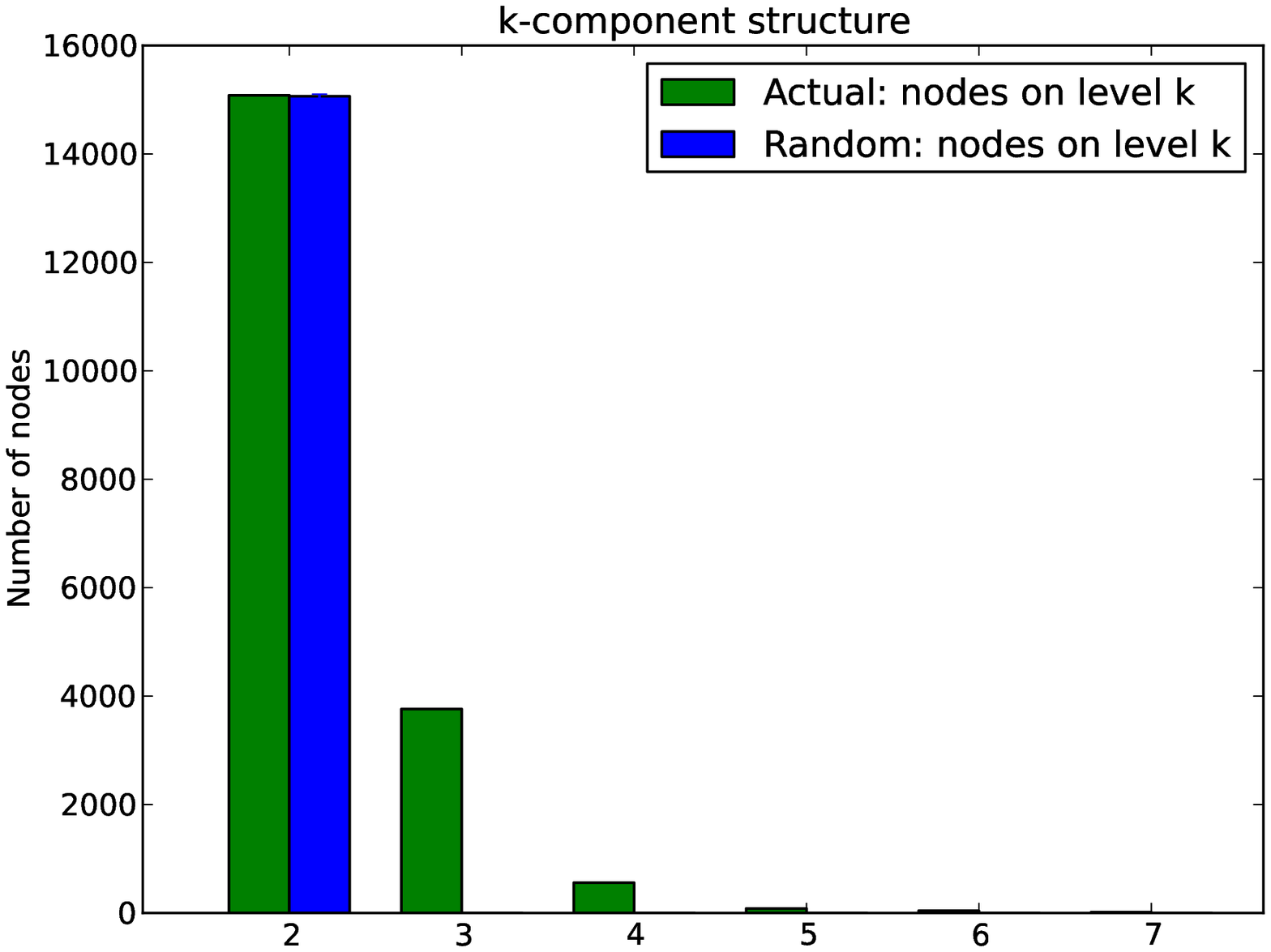}
}
\hspace{.05in}
\subfloat[Unipartite network formed by scientists during 5 years (2006-2010) in the high energy physics (theory) section of arXiv.org]{
\label{fig:hep_th_1m}
\includegraphics[scale=0.30]{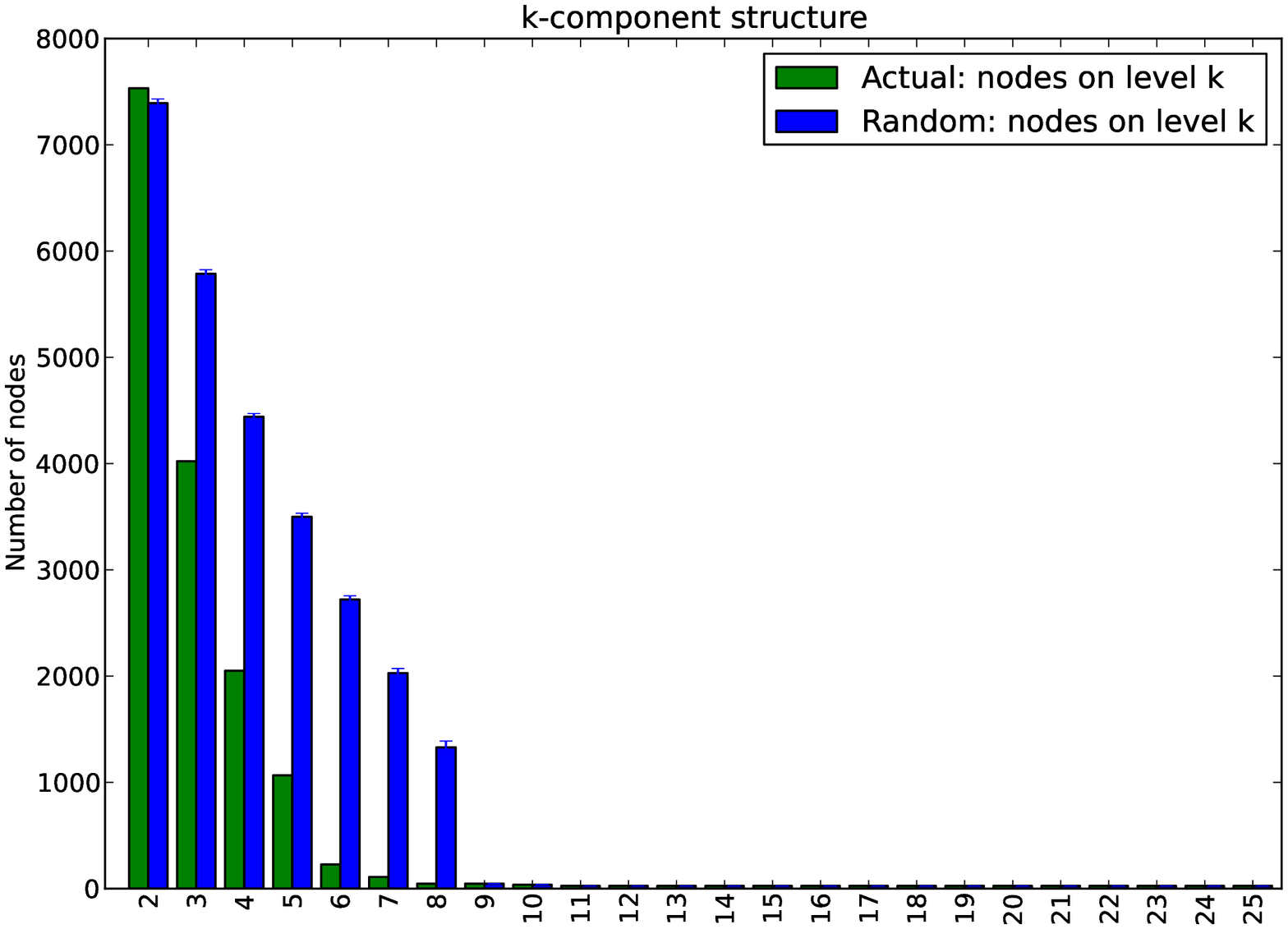}
}
\hspace{.01in}
\subfloat[Bipartite network formed by scientists and preprints during 5 years (2006-2010) in the nuclear physics (theory) section of arXiv.org]{
\label{fig:nucl_th_2m}
\includegraphics[scale=0.35]{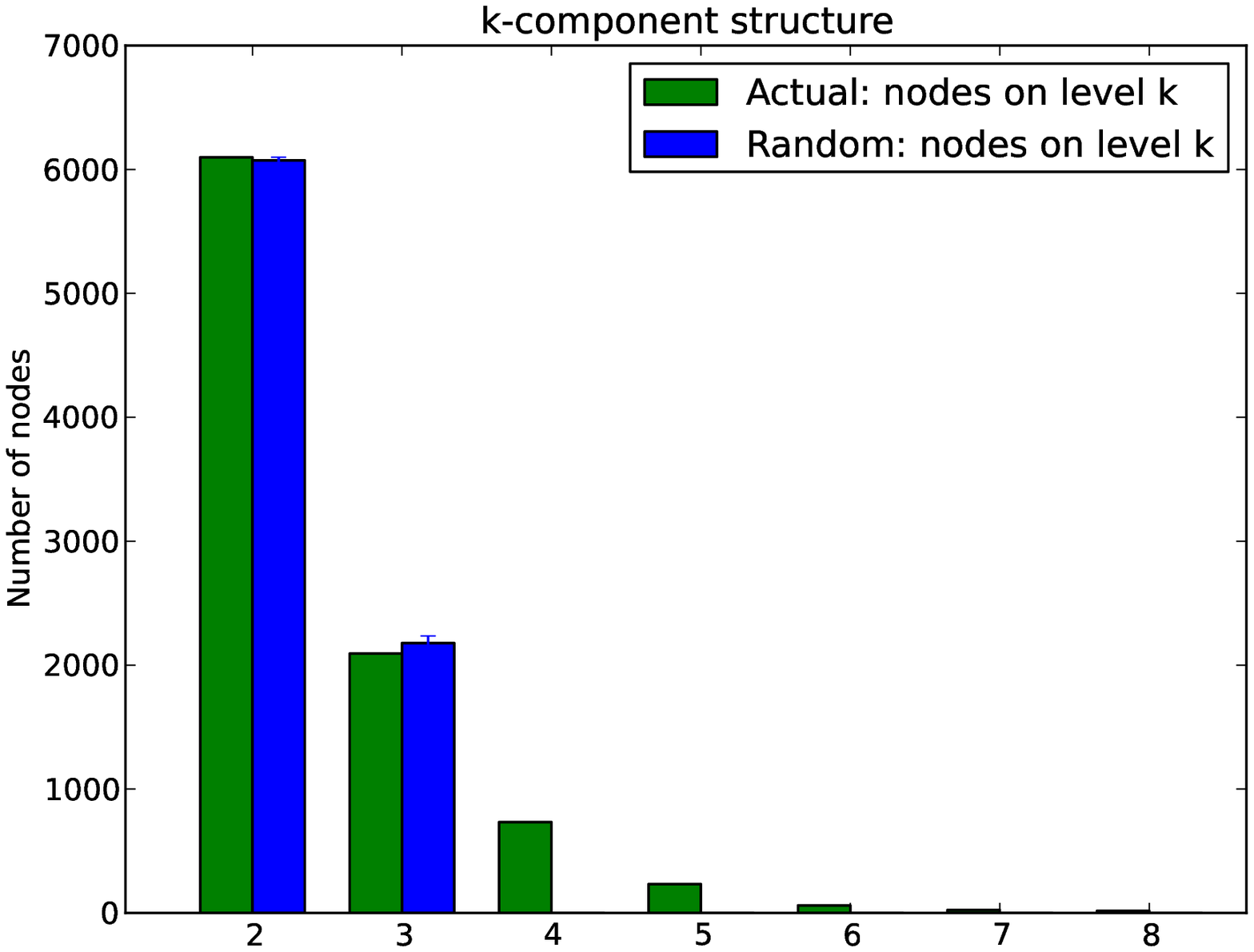}
}
\hspace{.05in}
\subfloat[Unipartite network formed by scientists during 5 years (2006-2010) in the nuclear theory section of arXiv.org]{
\label{fig:nucl_th_1m}
\includegraphics[scale=0.27]{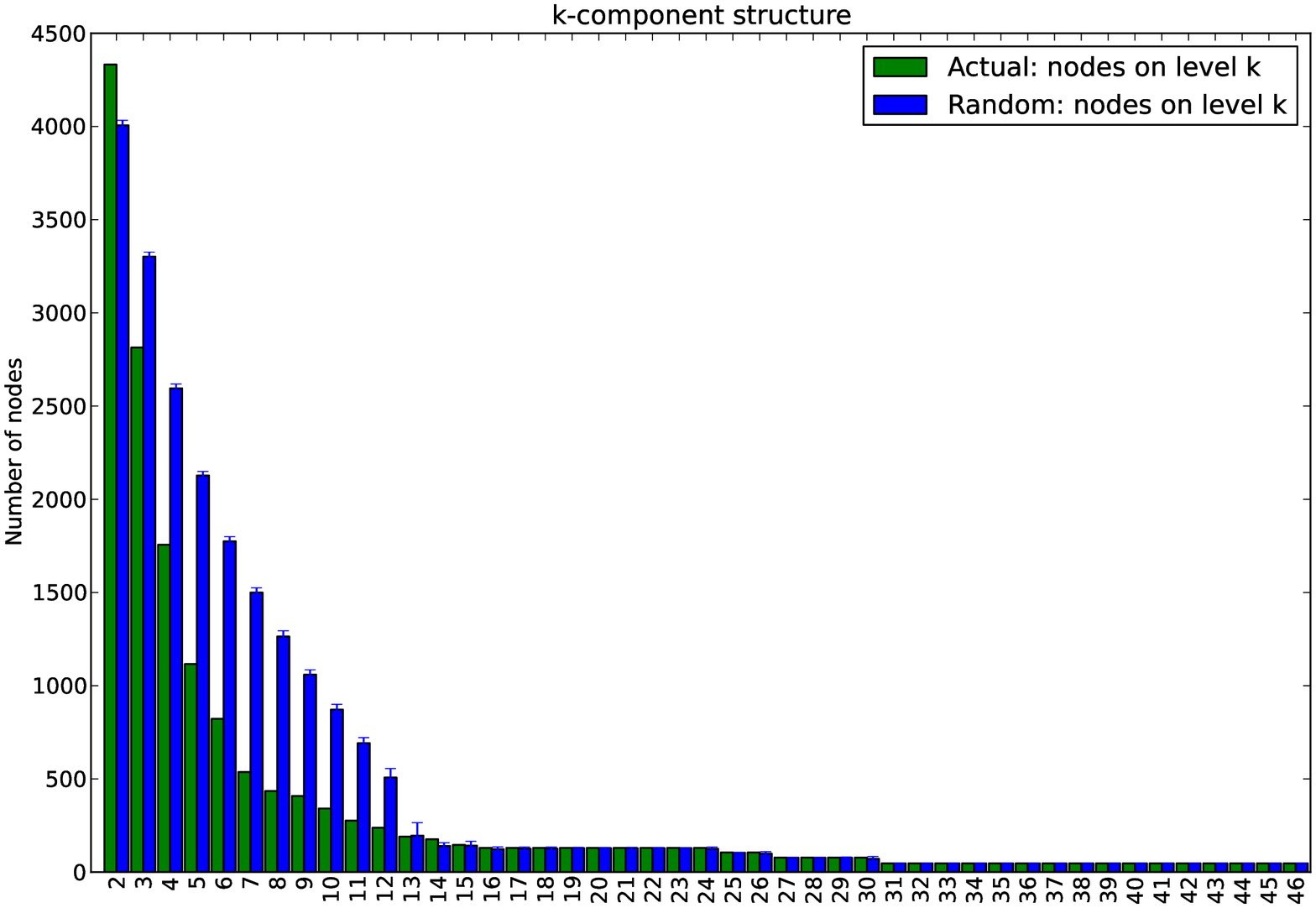}
}
\caption{Barplots of $k$-number frequencies for two-mode and one-mode collaboration networks and their random counterparts. Green bars represent the actual $k$-number frequencies and blue bars represent the average $k$-number frequencies for 64 random networks that maintain the degree distribution of the original two-mode network.} 
\label{fig:sc}
\end{figure}

Note that, so far, the comparison of actual networks with their random counterparts has focused on a single random network. But, a single random network is not a sound null model. We do need to generate a large enough set of them and perform the connectivity analysis to have an accurate picture of possible connectivity structures generated solely by chance given the observed bipartite degree distribution. A good way to evaluate the differences between the actual network and the set of random networks is comparing the frequencies of $k$-numbers of their nodes. A node's $k$-number, or component number, is the value $k$ of the highest order $k$-component in which it is embedded. In the barplots displayed in figure \ref{fig:sc}, each bar represents the number of nodes that have $k$-number $k$. Green bars represent $k$-number frequencies for the actual networks and blue bars represent the average value of 64 random networks that maintain the degree distribution of the original two-mode network. We analyzed 64 random networks to keep computation time reasonable, but we generated ten times more random networks and we have randomly selected one of each ten to perform the actual analysis.

Figure \ref{fig:sc} shows that two-mode and one-mode projections of the same network yield quite different results in terms of $k$-number distribution among nodes when compared with their random counterparts. Bipartite collaboration networks have slightly fewer nodes with low component number (2 and sometimes 3) than their random counterparts. However, they have a lot more nodes in higher levels of connectivity. This means that, in bipartite random networks, the edges are more evenly distributed among all nodes. Thus more nodes are embedded in bicomponents, and in some cases, tricomponents; but also for this same reason, random networks have a lot fewer nodes in $k$-components of higher order (4, 5 or 6) than actual networks. Therefore, we can conclude that bipartite collaboration networks are significantly more hierarchical in connectivity terms than their random counterparts. As this hierarchy cannot be explained in terms of random mixing papers/programs with scientists/developers, it must be the result of an underlying organization principle that shapes the structure of these collaboration networks.

Going one step beyond classical structural cohesion analysis, as proposed above, we can deepen our analysis by also considering the average connectivity of the $k$-components of these networks. By analogy with the $k$-component number of each node, which is the maximum value $k$ of the deepest $k$-component in which that node is embedded, we can establish the average $k$-component number of each node as the value of average connectivity of the deepest $k$-component in which that node is embedded. Notice that, unlike plain node connectivity, average node connectivity is a continuous measure of cohesion. Thus it provides a more granular measure of cohesion because we can rank $k$-components with the same $k$ according to their average node connectivity.

\begin{figure}[p]
%\centering
\subfloat[Debian Lenny 2 mode]{
\label{fig:s3d_lenny_2m}
\includegraphics[scale=0.35]{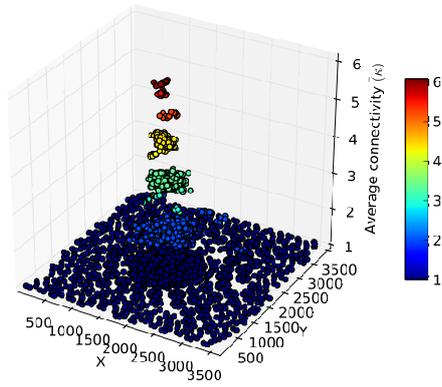}
}
\hspace{.01in}
\subfloat[Debian Lenny 1mode]{
\label{fig:s3d_lenny_1m}
\includegraphics[scale=0.35]{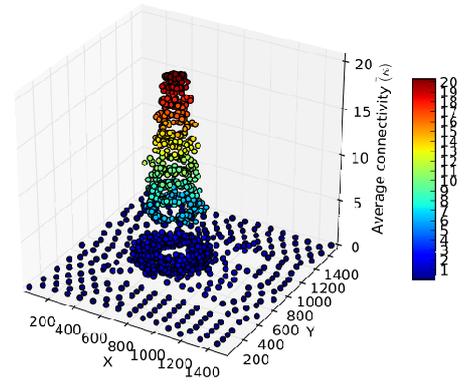}
}

\subfloat[Nuclear Theory 2 mode]{
\label{fig:s3d_nucl_2m}
\includegraphics[scale=0.35]{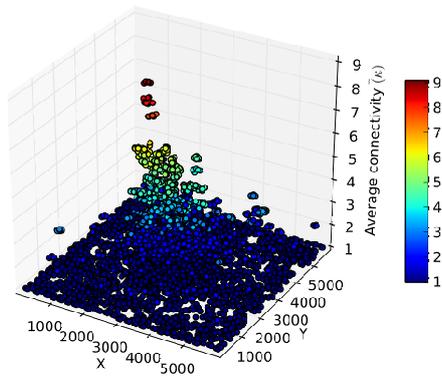}
}
\hspace{.01in}
\subfloat[Nuclear Theory 1 mode]{
\label{fig:s3d_nucl_1m}
\includegraphics[scale=0.35]{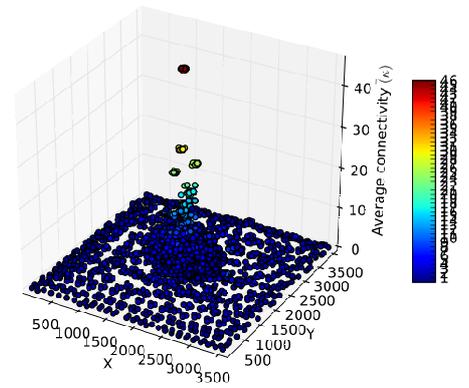}
}

\subfloat[High Energy Theory 2 mode]{
\label{fig:s3d_hep_2m}
\includegraphics[scale=0.35]{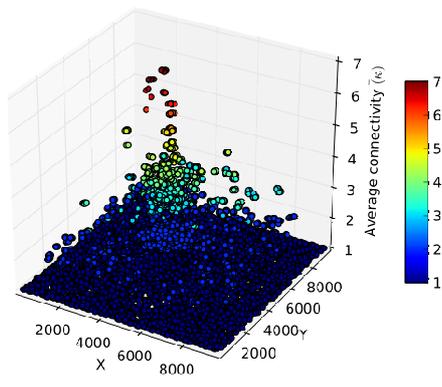}
}
\hspace{.01in}
%\subfloat[High Energy Theory 1 mode]{
%\label{fig:s3d_hep_1m}
%\includegraphics[scale=0.35]{scatter_3d_kk_hep_th_1mode}
%}

\caption{Average connectivity three-dimensional scatter plots. X and Y are the positions determined by the Kamada-Kawai layout algorithm. The vertical dimension is average connectivity. Each dot is a node of the network and two-mode networks contain both papers/programs and scientists/developers.}
\label{fig:scatter3d}
\end{figure}

Figure \ref{fig:scatter3d} graphically represents the three networks with three-dimensional scatter plots\footnote{These plots are produced with the powerful Matplotlib python library \citep{hunter:2007}.}.  In these graphs, each dot corresponds to a node of the network, for two-mode networks nodes represent both scientists/developers and papers/programs. The Z axis (the vertical one) is the average $k$-component number of each node, and the X and Y axis are the result of a 2 dimensional force-based layout algorithm implemented by the \texttt{neato} program of Graphviz \citep{graphviz}. The two dimensional layout is computed by constructing a virtual physical model and then using an iterative solver procedure to obtain a low-energy configuration. Following \citet{kamada:1989}, an ideal spring is placed between each pair of nodes (even if they are not connected in the network). The length of each spring corresponds to the geodesic distance between the pair of nodes that it links. The final node positioning in the layout approximates the path distance among pairs of nodes in the network.

This novel graphic representation of cohesion structure is inspired by the approximation technique developed by \citet{moody:2004} for plotting the approximate cohesion contour of large networks to which is not practical to apply \citeauthor{moody:2003}'s exact algorithm for $k$-components \citeyear{moody:2003}. Moody's technique is based on the fact that force-based layouts algorithms tend to draw nodes within highly cohesive subgroups near each other. Then we have to divide the surface of the two-dimensional plane in squares of equal areas and compute node independent paths on a sample of pairs of nodes inside each square so as to obtain an approximation for the node connectivity in that square. Then we can draw a surface plot using a smoothing probability density function. However, in order to obtain a nice smooth surface plot, we have to use heavy smoothing in the probability density function, and carefully choose the area of the squares (mostly by trial and error). Moreover, this technique strongly relies on the force-based layout algorithm to put nodes in highly cohesive subgroups near each other ---something which is not guaranteed because they are usually based in path distance and not directly on node connectivity. Because we are able to compute the $k$-component structure with our heuristics for large networks, the three-dimensional scatter plot only relies on the layout algorithm for setting the X and Y positions of the nodes, while the Z position (average node connectivity) is computed directly from the network. Moreover, we don't have to use a smoothed surface plot because we have a value of average connectivity for each node, and thus we can plot each node as a dot on the plot. This gives a more accurate picture of the actual cohesive structure of a network.

Our synthetic representation of their cohesive structures can help researchers visualize the presence of different organizational mechanisms in different kinds of collaboration networks. The difference between the Debian and the scientific collaboration networks is striking. In figure \ref{fig:s3d_lenny_2m} we can see the scatter plot for a Debian bipartite network. We can observe a clear vertical separation among nodes in different connectivity levels. This is because almost all nodes in each connectivity level are in a giant $k$-component and thus they have the same average connectivity. In other words, developers in Debian show different levels of engagement and contribution, with a core group of developers deeply nested at the core of the community.  This pattern is the result of formal and informal rules of collaboration that evolved over the years \citep{ferraro:2007} into a homogeneous hierarchical structure, where there is only one core of highly productive individuals at the center. Not surprisingly, perhaps, the Debian project has been particularly resilient to developers' turnover and splintering factions.

Scientific collaboration networks show a rather different structure of collaboration. The two-mode science collaboration networks (figures \ref{fig:s3d_nucl_2m} and \ref{fig:s3d_hep_2m}) display a continuous hierarchical structure in which there are nodes at different levels of average connectivity for each discrete plain connectivity level. This is because science collaboration networks have a complex cohesive block structure where there are a lot of independent $k$-components in each plain connectivity level, for $k \ge 3$. Each small cohesive block has a different order, size and average connectivity; thus, when we display them in this three-dimensional scatter plot we observe a continuous hierarchical structure that contrasts with the almost discrete structure of Debian collaboration networks.

One explanation why we observe this heterogeneous connectivity structure is that scientific collaborations cluster around a variety of different aims, methods, projects, and institutional environments. Therefore as the most productive scientists collaborate with each other, hierarchies naturally emerge. However, we are less likely to observe one single hierarchical order as we did in the Debian network, as more than one core of highly productive scientists is likely to emerge. In a way our visualization captures the structure of the ``invisible college'' of the scientific discipline.

If we compare the bipartite networks with their one-mode projections using this graphical representation (see figures \ref{fig:s3d_lenny_1m}, \ref{fig:s3d_nucl_1m}, and \ref{fig:s3d_hep_1m}) we can see that, again, they look quite different. While bipartite average connectivity structure for the Debian network is characterized by clearly defined and almost discrete hierarchical levels, its one-mode counterpart shows a continuous hierarchical structure. However, this is not caused by the presence of many small $k$-components at the same level $k$, as in the case of bipartite science networks discussed above, but by the close succession of hierarchy levels with almost the same number of nodes in a chain-like structure (as depicted in figure \ref{fig:cb_lenny_1m}).

For collaboration science networks, the three-dimensional scatter plots of one-mode projections are also quite different than their original bipartite networks. They have a lot more hierarchy levels than bipartite networks but most nodes are at lower connectivity levels. Only a few nodes are at top levels of connectivity, and they all form part of some clique, which are the groups in the long ``legs'' of the cohesive block structure depicted in figure \ref{fig:cb_nucl_1m}. Thus, the complex hierarchical connectivity structure of bipartite collaboration networks gets blurred when we perform one-mode projection. An important consequence of the projection is that only a few nodes embedded in big cliques appear at top connectivity levels and all other nodes are way down in the connectivity structure. This could lead the risk of overestimating the importance of those nodes in big cliques and to underestimate the importance of nodes that, despite being at high levels of the bipartite connectivity structure, appear only at lower levels of the unipartite connectivity structure.

\section{Conclusions}

This article contributes to our understanding of structural cohesion in a number of ways.

First, we extended theoretically the structural cohesion model by considering not only plain node connectivity, which is the minimum number of nodes that must be removed in order to disconnect a network, but also the average node connectivity of networks and its cohesive groups, which is the number of nodes that, on average, must be removed to disconnect an arbitrary pair of nodes in the network. Taking into account average connectivity allows a more granular conception of structural cohesion, and we show in our empirical analysis of collaboration networks how this approach leads to useful implications in empirical research.

Second, we developed heuristics to compute the $k$-components structure, along with the average node connectivity for each $k$-component, based on the fast approximation to compute node independent paths \citep{white:2001b}. These heuristics allow for the computing of the approximate value of group cohesion for moderately large networks, along with all the hierarchical structure of connectivity levels, in a reasonable time frame. We showed that these heuristics can be applied to networks at least one order of magnitude bigger than the ones manageable by the exact algorithm proposed by \citet{moody:2003}. To ensure reproducibility and facilitate diffusion of these heuristics we provided a very detailed description of the implementation, along with an illustration of the source code \footnote{We believe that providing detailed implementation is critical to ensure reproducibility, but often these details are black-boxed, some times because of proprietary software restrictions or authors' reluctance to share their work.}. 

Finally, we used the heuristics proposed here to analyze three large collaboration networks. With this analysis, we showed that the heuristics and the novel visualization technique for cohesive network structure help us capture important differences in the way collaboration is structured. Obviously a detailed analysis of the institutional and organizational structures in which the collaborative activity took place is well beyond the scope and aims of this paper. But future research could leverage the tools we provide to systematically measure those structures. For instance, sociologists of science often compare scientific disciplines in terms of their collaborative structures \citep{moody:2004} and their level of controversies \citep{bearman:2010}. The measures and the visualization technique we proposed could nicely capture these features and compare them across scientific disciplines. This would make it possible to further our understanding of the social structure of science, and its impact in terms of productivity, novelty and impact. Social network researchers interested in organizational robustness would also benefit from leveraging the structural cohesion measures to detect sub-groups that are more critical to the organization's resilience, and thus prevent factionalization. Exploring the consequences of different forms of cohesive structures will eventually help us further our theoretical understanding of collaboration and the role that cohesive groups play in linking micro-level dynamics with macro-level social structures.

\newpage

\appendix
\gdef\thesection{Appendix \Alph{section}}

\section{Illustration of the heuristics}
\label{illustration}

\begin{figure}[h!]
\centering
%\subfloat[Nodes colored by core number. Notice that the 3-core includes all nodes in the graph.]{
%\label{fig:core}
%\includegraphics[scale=0.32]{illustrative_core}
%}
%\hspace{.05in}
\subfloat[Nodes colored by component number according to our algorithm. Note the error when two $K_5$ overlap in two nodes]{
\label{fig:component}
\includegraphics[scale=0.36]{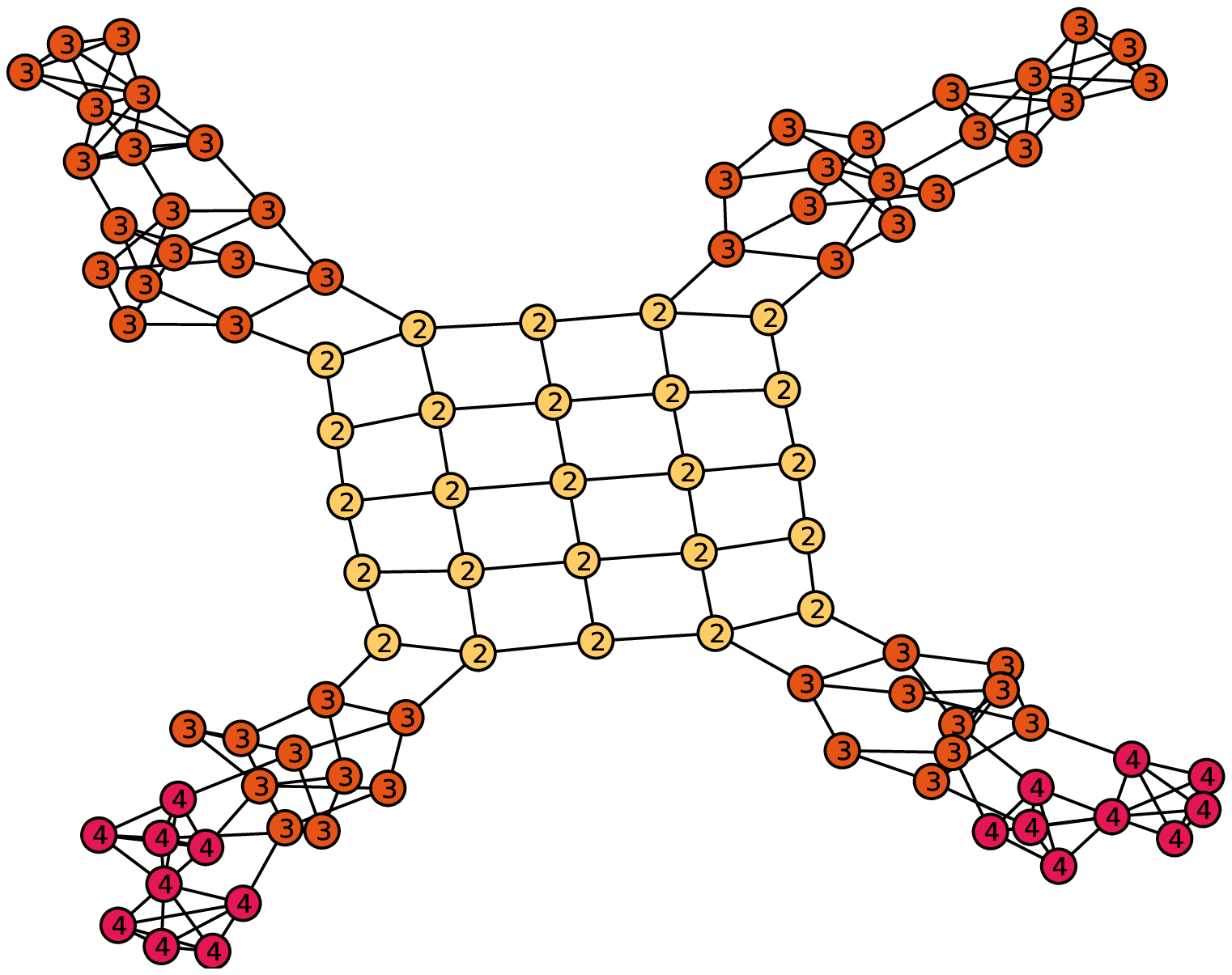}
}
\hspace{.05in}
\subfloat[Nodes colored by component number according to Moody \& White algorithm.]{
\label{fig:component}
\includegraphics[scale=0.36]{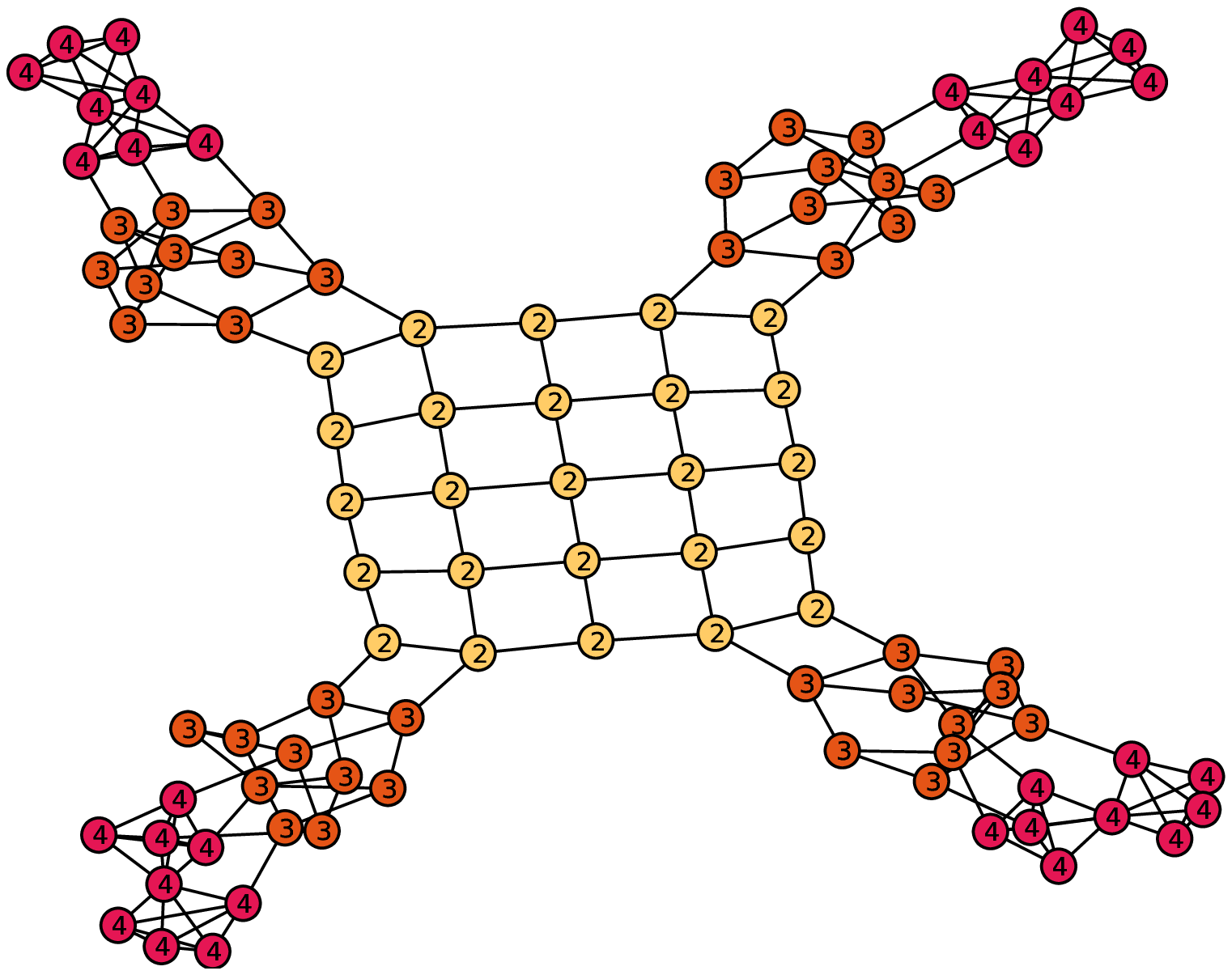}
}
\hspace{.05in}

\caption{Synthetic graph composed of a two dimensional grid of 25 nodes, four Petersen graphs ($P$) with ten nodes each (with $\kappa = 3$) linked by two edges to the grid, and eight complete graphs $K_5$ (with $\kappa = 4$) linked by three edges to each Petersen graph. In two cases $K_5$ overlap in 1 node and in the other two cases they overlap in 2 nodes. The whole graph is biconnected and also a tricore. Notice that our algorithm fails to classify the two $K_5$ that overlap in two nodes as 4-components. See text and  figure figure  \ref{fig:example_4} for details.}
\label{fig:example}
\end{figure}

In order to illustrate how the proposed heuristics works, we will use a convenient synthetic network with 99 nodes and 200 edges where $\kappa \neq \delta$. This network is based on a two dimensional grid of 5 by 5 nodes. In each corner of the grid we attach a Petersen graph ($P$), linked by two edges to the grid. Thus the only four nodes of the grid with degree 2 are linked to a Petersen graph. All nodes of the grid are therefore part of a 3-core. Each $P$ is linked to two complete graphs with 5 nodes ($K_5$); in two cases those two $K_5$ overlap in only one node and in the other two cases, they overlap in two nodes. The Petersen graph is linked by three edges to one of the $K_5$, thus making one of each $K_5$ part of a tricomponent along with $P$. In the case of the two $K_5$ that overlap only on one node, the outer $K_5$ has also one edge linking one of its nodes with one node of $P$ nodes, in order to make the whole graph biconnected (see figure \ref{fig:example}). Petersen graphs have node connectivity 3 and complete graphs with 5 nodes have node connectivity 4. Notice that the whole example graph is biconnected and a 3-core, but it has three levels of node connectivity: 2 for the grid, 3 for the Petersen graphs ($P$) and 4 for the complete graphs of 5 nodes ($K_5$).

%\begin{landscape}

\begin{figure}[h!]
\centering
\subfloat[Auxiliary graph $H$ for $k=3$ computed using White \& Newman's approximation algorithm for local node connectivity.]{
\label{fig:aux3}
\includegraphics[scale=0.36]{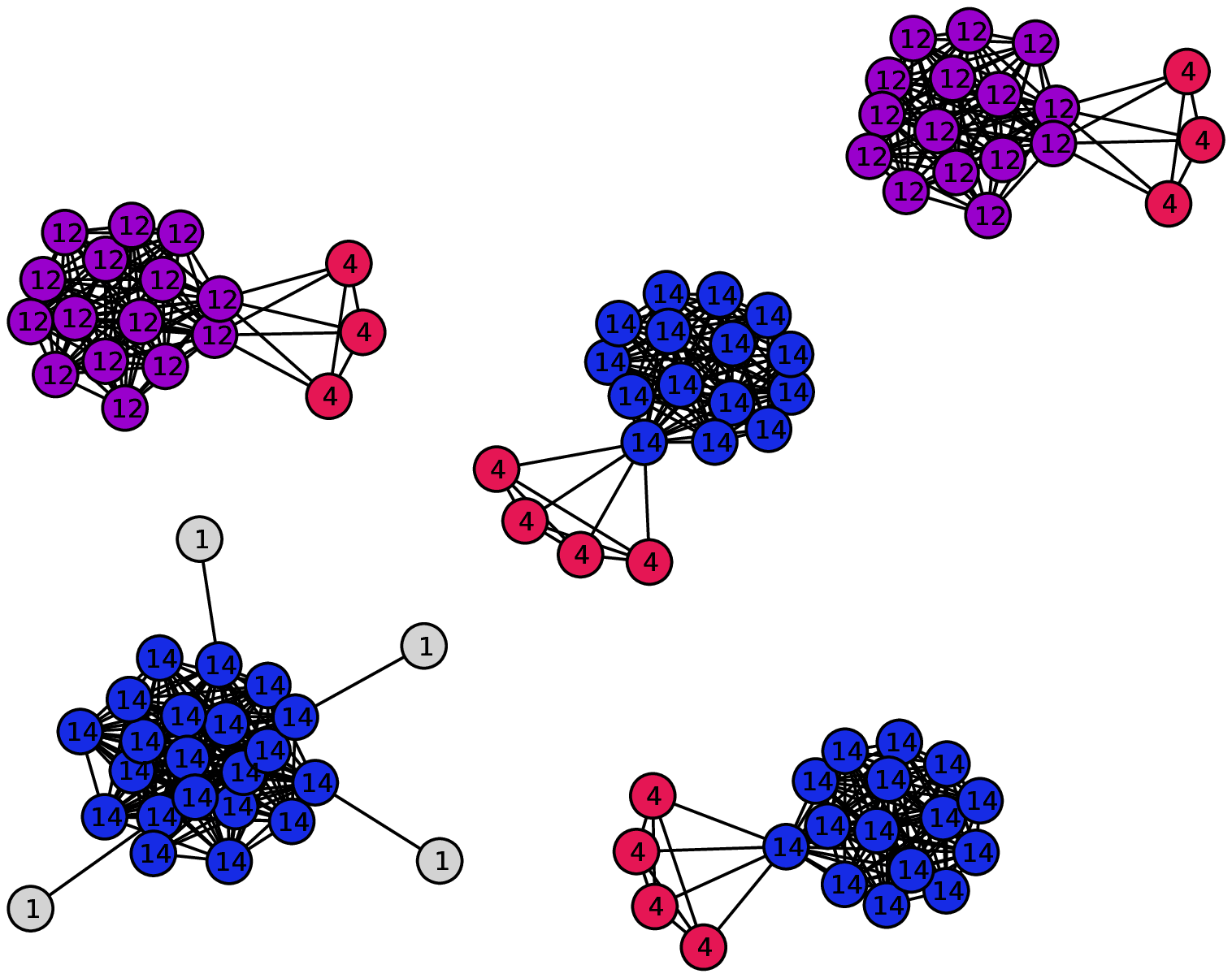}
}
\hspace{.05in}
\subfloat[Auxiliary graph $H$ for $k=3$ computed using flow-based connectivity algorithm for local node connectivity.]{
\label{fig:aux3_ex}
\includegraphics[scale=0.36]{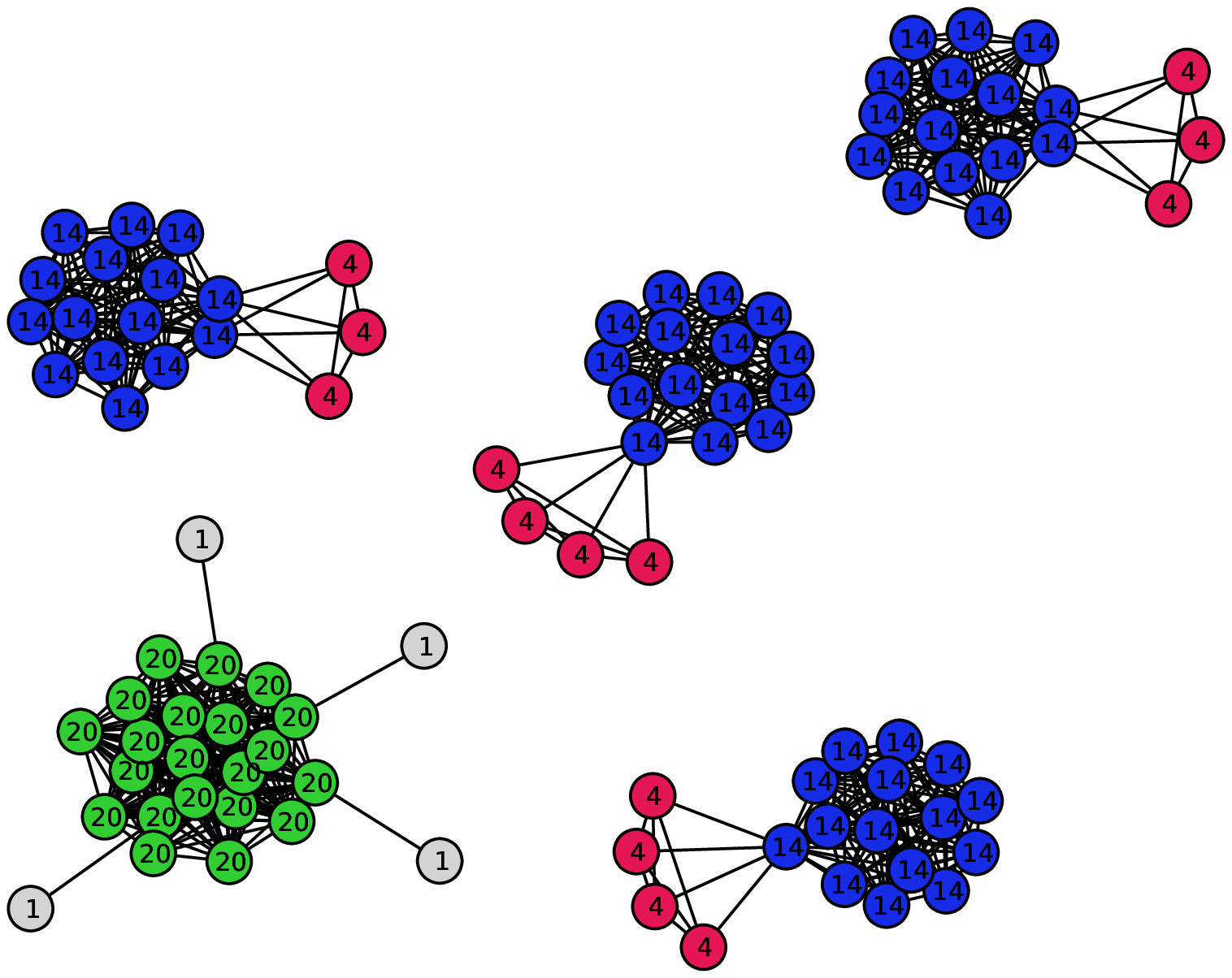}
}
\hspace{.05in}
\subfloat[All subgraphs $H_{candidate}$ from $H_3$ computed using White \& Newman's approximation algorithm for local node connectivity.]{
\label{fig:aux3_cand}
\includegraphics[scale=0.36]{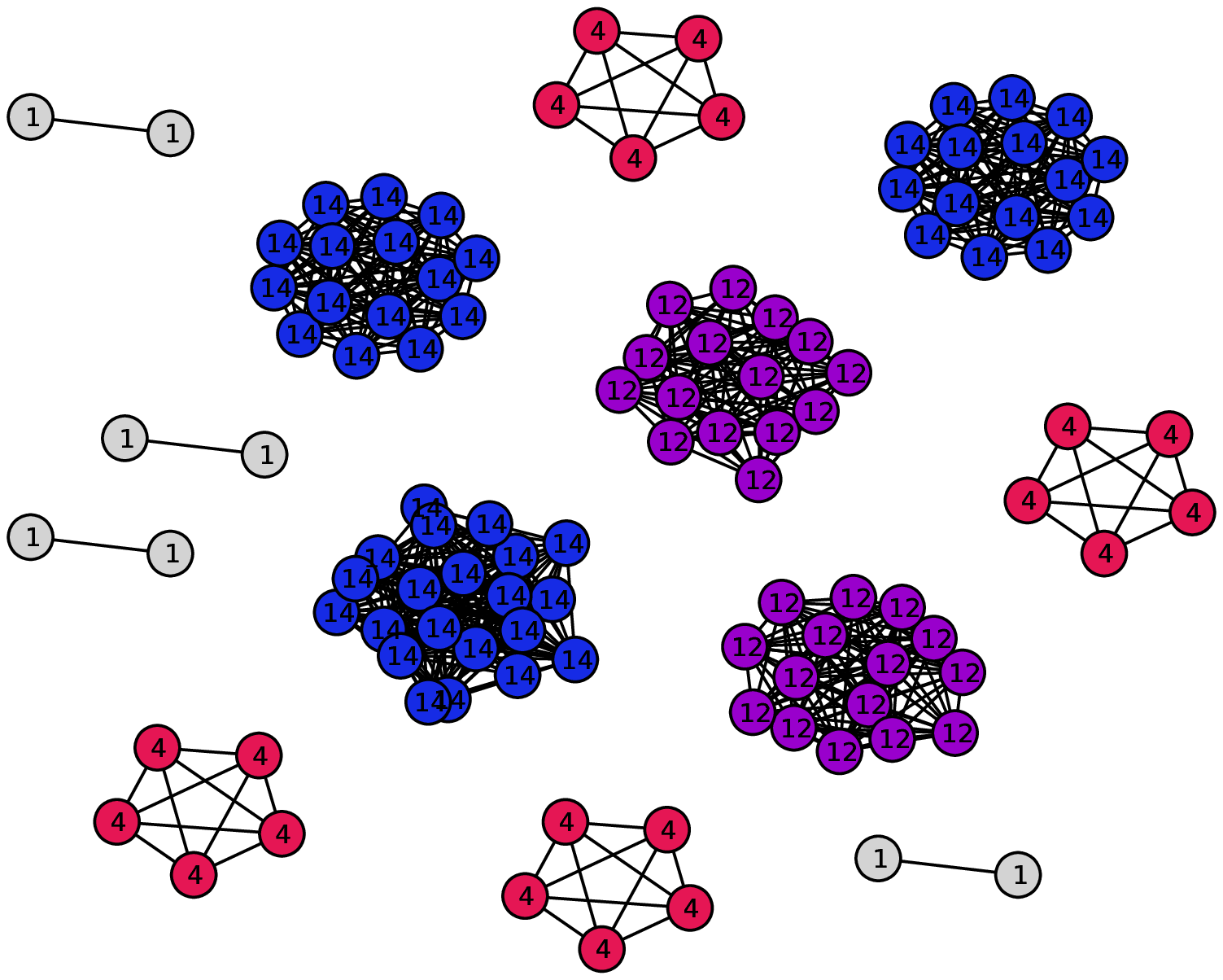}
}
\hspace{.05in}
\subfloat[All subgraphs $H_{candidate}$ from $H_3$ computed using flow-based connectivity algorithm for local node connectivity.]{
\label{fig:aux3_cand_ex}
\includegraphics[scale=0.36]{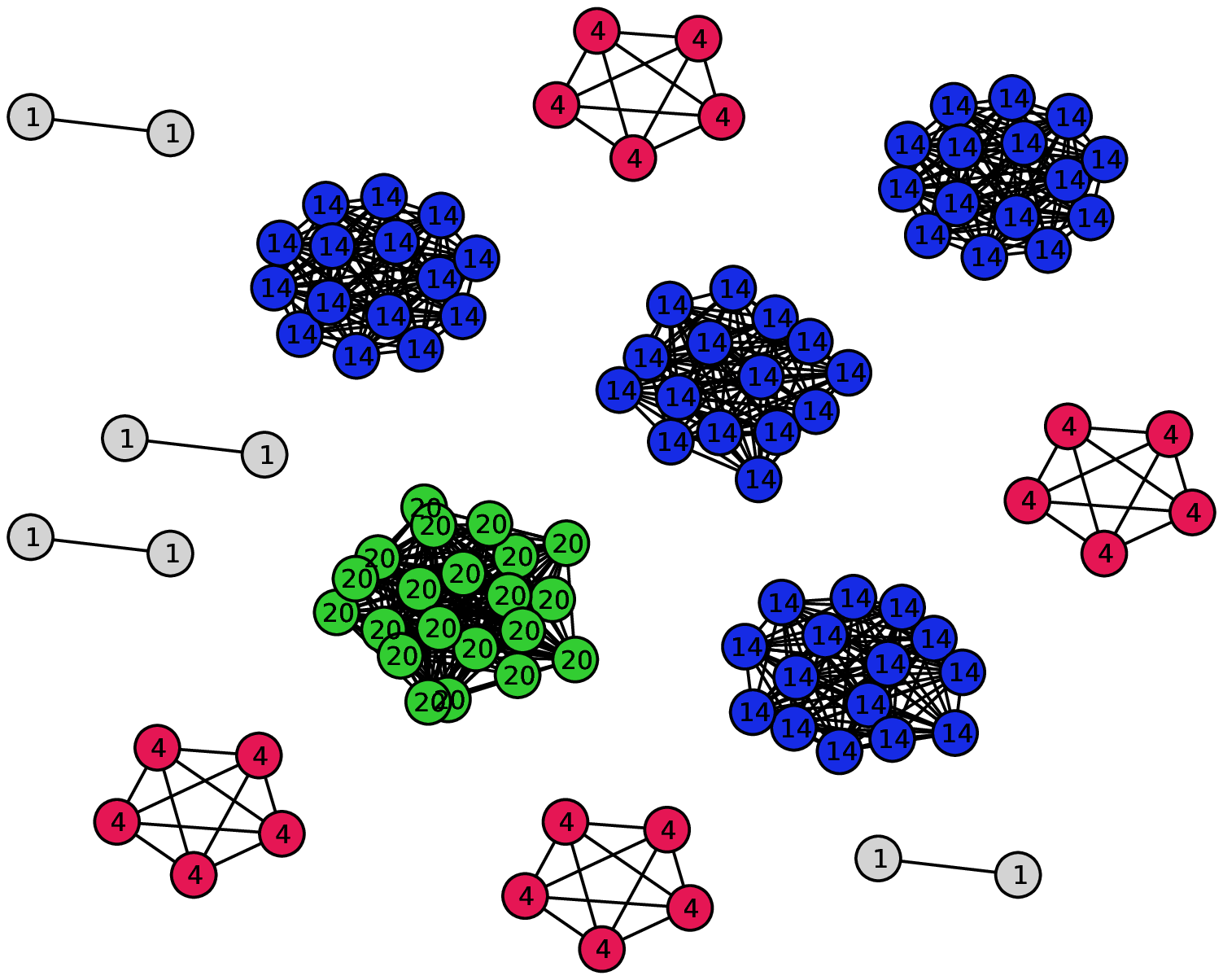}
}
\hspace{.05in}
\subfloat[Detected tri-components using the heuristics with the relaxation criteria of density $\ge 0.95$ in $H_{candidate}$.]{
\label{fig:subgraphs3}
\includegraphics[scale=0.36]{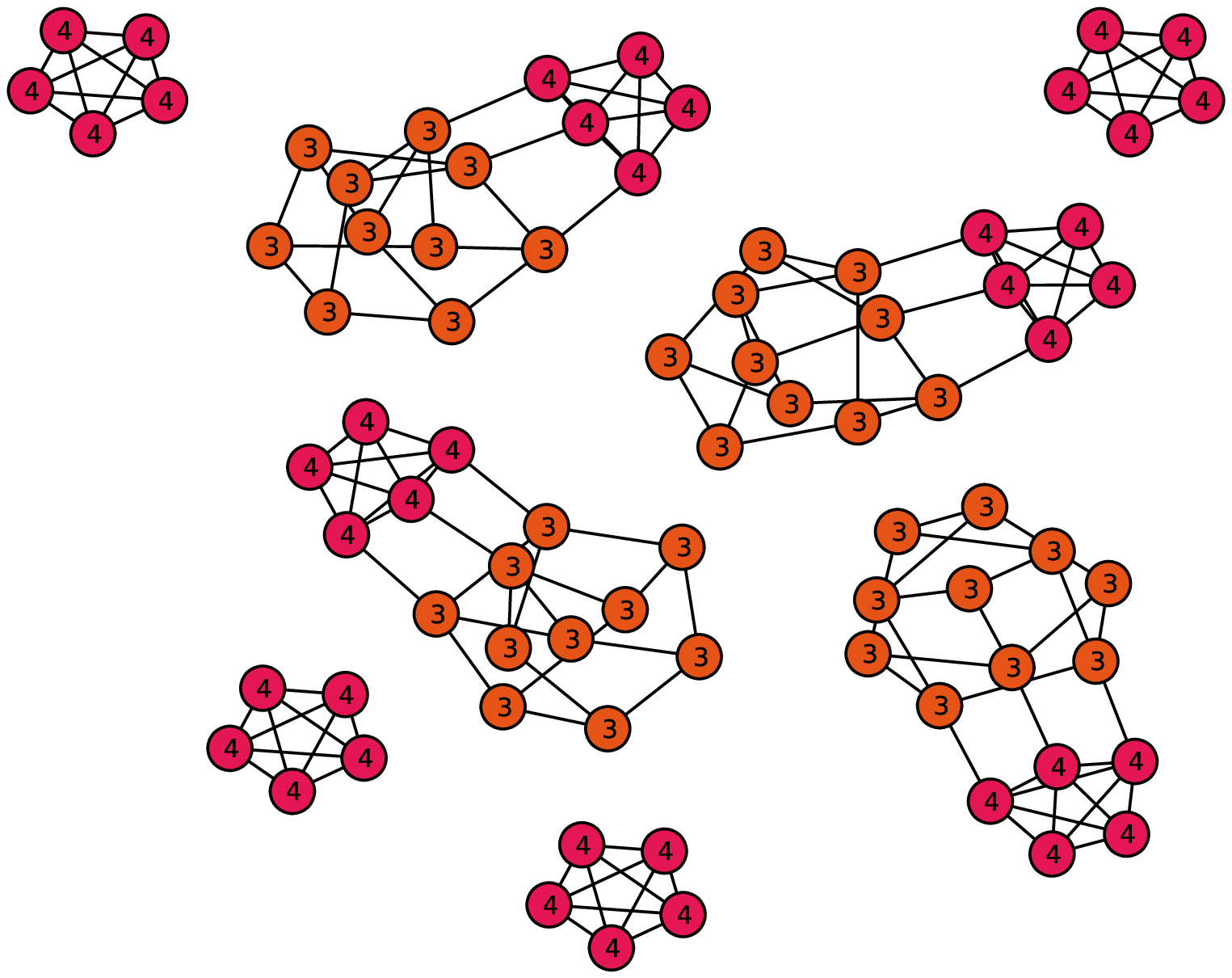}
}
\hspace{.05in}

\caption{Auxiliary graph $H_3$ for $k=3$. Note that when using White and Newman's approximation algorithm for local node connectivity (subfigure a), some node independent paths are not detected: the $P$ subgraphs linked to the two $K_5$ that overlap in two nodes should have core number 14 (blue) as in subfigure b, but they have core number 12. Thus to correctly detect all tricomponents we have to set a relaxation criteria for $H_{candidate}$, in this example setting density at 0.95 or allowing a variation of 2 in the degree of all nodes of $H_{candidate}$, allows the algorithm to correctly detect all tricomponents.}
\label{fig:example_3}
\end{figure}

%\end{landscape}

\begin{figure}[h!]
\centering
\subfloat[Auxiliary graph $H$ for $k=4$ computed using White and Newman's approximation.]{
\label{fig:aux4}
\includegraphics[scale=0.36]{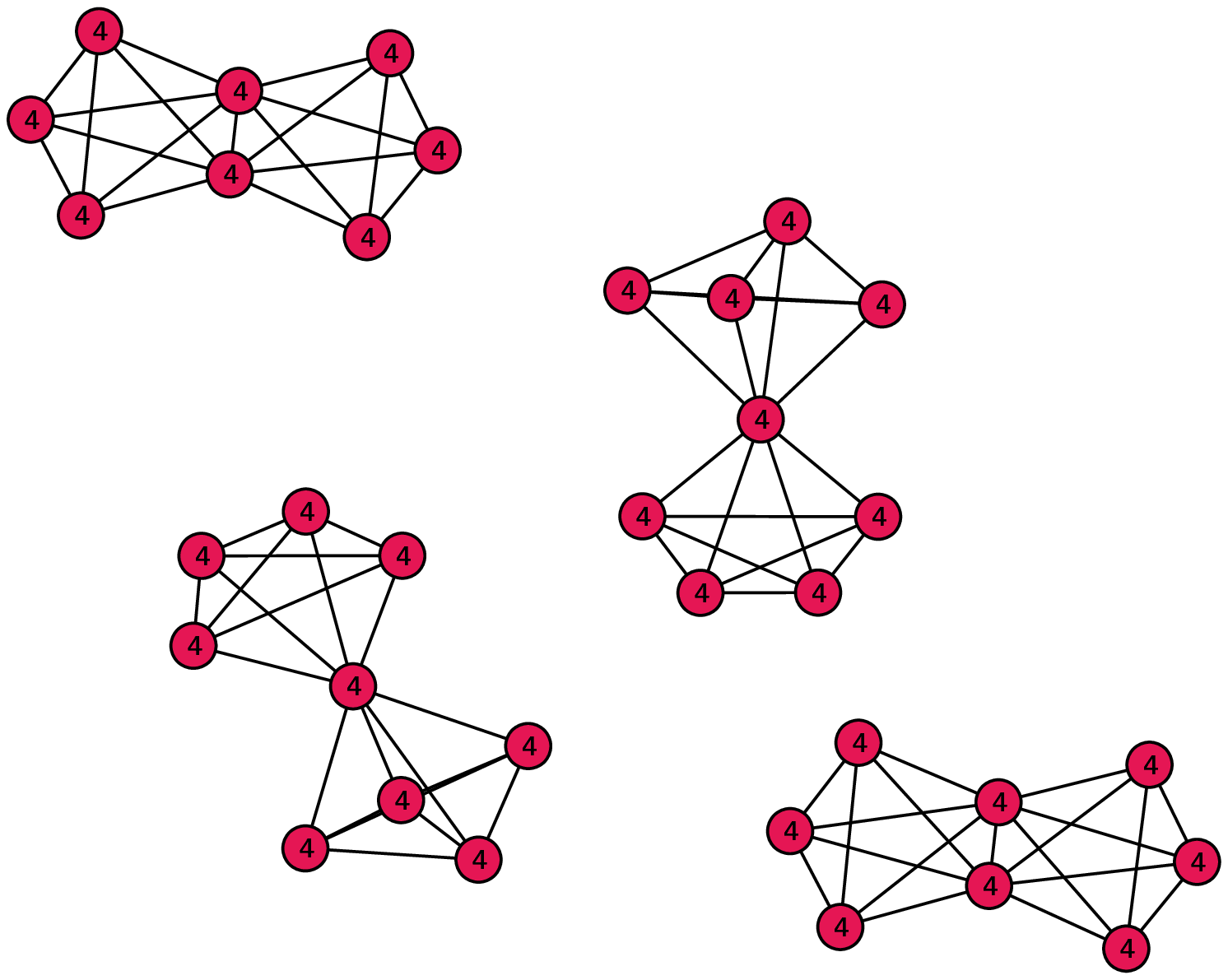}
}
\hspace{.05in}
\subfloat[Detected 4-components using our heuristics. Note that there should be four more $K_5$, the ones that overlap in two nodes are not detected as 4-components. See text for an explanation.]{
\label{fig:subgraphs4}
\includegraphics[scale=0.36]{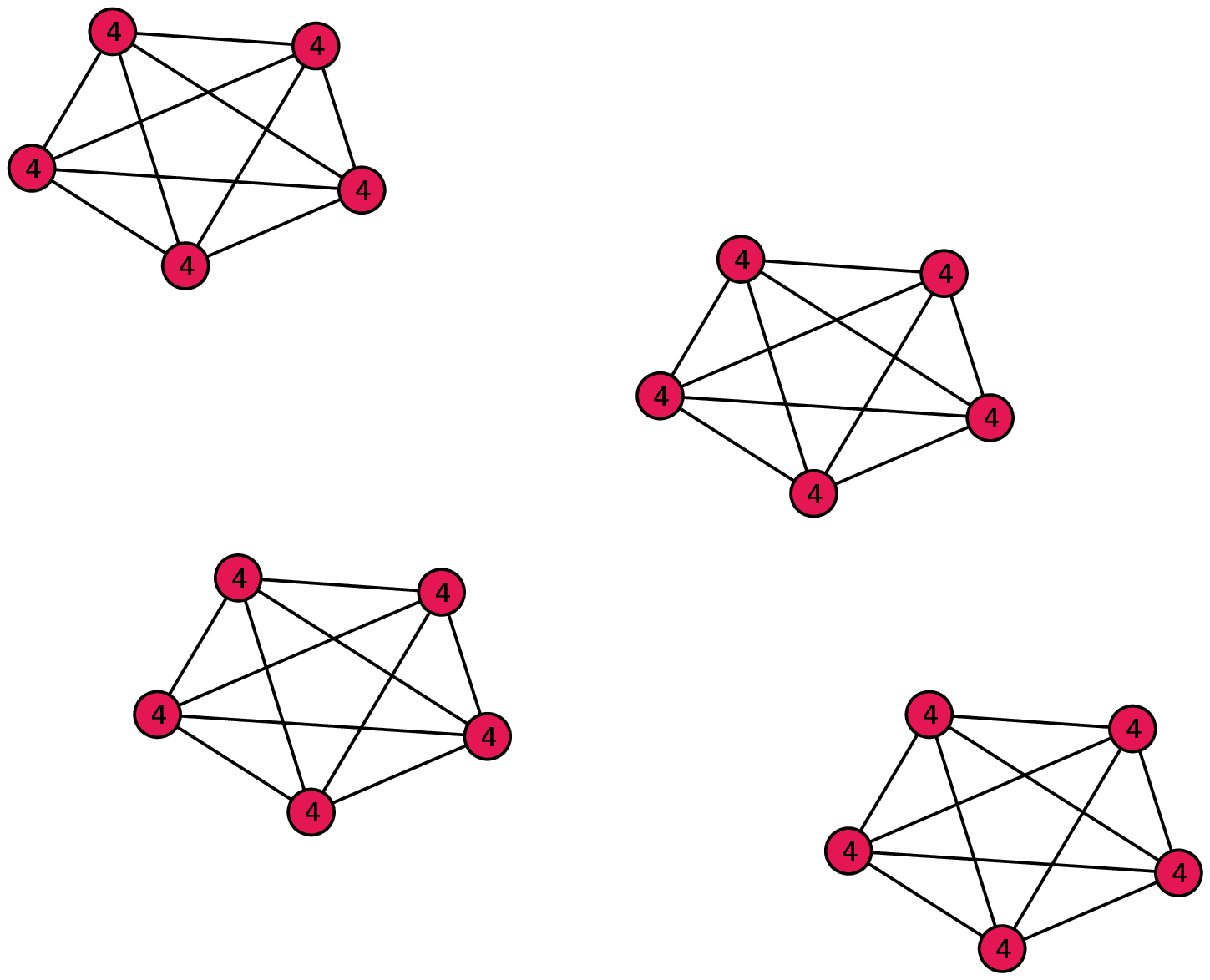}
}

\caption{Auxiliary graph $H_4$ for $k=4$. In this case both White and Newman's approximation algorithm, and the exact flow-based algorithm for local node connectivity yield equal results. Note that there should be four more $K_5$ in subfigure b, the ones that overlap in two nodes are not detected as 4-components. This is because, as can be seen in subfigure a, the nodes in these $H_{candidate}$ subgraphs have all the same core number, but their density is 0.67 and the difference in degree is 3. Thus, in order to detect them we would have to relax the clique criteria for $H_{candidate}$ too much, and even then we would classify both $K_5$ as a single 4-component, which is obviously wrong.}
\label{fig:example_4}
\end{figure}

As discussed above, a $k$-core is a maximal subgraph that contains nodes of degree $k$ or more. The core number of a node is the largest value $k$ of a $k$-core containing that node. On the other hand, a $k$-component is a maximal subgraph that cannot be disconnected by removing less than $k$ nodes. The component number of a node is the largest value $k$ of a $k$-component containing that node.

The graph of figure \ref{fig:example} is a biconnected 3-core, which means that it is a graph with minimum degree $=3$ that cannot be disconnected by removing less than 2 nodes. Our algorithm starts by considering the whole graph the step 2, but in $k$-core subgraphs with more than one bicomponent, the following steps are performed for each bicomponent of the $k$-core. We will only compute up until $k=4$ because the largest core number of a node in $G$ is 4. 

For $k=3$ we create an auxiliary graph with all biconnected nodes with core number $\ge 3$ (see figure \ref{fig:example_3}). In this case all nodes have a core number greater than or equal to 3. Thus the auxiliary graph $H$ for $k=3$ contains all 99 nodes. We then link two nodes in $H_3$ if we can find $k$ or more node independent paths between them. As we can see, the result are five connected components, four of which correspond to each Petersen graph plus the two $K_5$, while the last one corresponds to the nodes that form the grid. The later has 4 nodes that are linked by 3 node independent paths to only one node, these four nodes are the four corner nodes of the grid. 

Notice that when using White and Newman's approximation algorithm for local node connectivity (subfigure \ref{fig:aux3}), some node independent paths that actually exist are not detected: the $P$ subgraphs linked to the two $K_5$ that overlap in two nodes should have a core number of 14 (blue) because there are 3 node independent paths linking each pair of different nodes in the subgraph formed by the $P$ and the $K_5$ to which it is linked through three edges, as in subfigure \ref{fig:aux3_ex}, which was computed using the exact flow-based algorithm for local node connectivity. Notice also that the grid has core number 14 in \ref{fig:aux3} but actually should be core number 20 as shown in \ref{fig:aux3_ex}. This illustrates the importance of computing biconnected components of $H$ (setp 3.c) before building the subgraphs $H_{candidate}$ (step 3.d).

Figures \ref{fig:aux3_cand} and \ref{fig:aux3_cand_ex} depict $H_{candidate}$ subgraphs, the former using White and Newman's approximation algorithm and the latter using an exact flow-based algorithm for local node connectivity. The subgraphs $H_{candidate}$ are composed by nodes that are in the same biconnected component of $H$ and have \emph{exactly} the same core number. Notice that in figure \ref{fig:aux3_cand} the $P$ graphs linked to the two $K_5$ that overlap in two nodes have core number $< n-1$ (the magenta clusters), thus they are not complete (density=0.96) and the degree of their nodes is not homogeneus: two nodes have degree 12, four have degree 13, and nine have degree 14. Therefore, if we enforce the clique critera for $H_{candidate}$ we would not detect all tricomponents because, following the algorithm, we would have to start removing nodes with the lowest degree and check if at some point we find a complete subgraph. In order to correctly detect all tricomponents in this illustrative example, we have to first establish a relaxation for the clique criteria for $H_{candidate}$. In this case, setting density at 0.95 or allowing a variation of 2 in the degree of all nodes of $H_{candidate}$, allows the algorithm to correctly detect all tricomponents as shown in figure \ref{fig:subgraphs3}.

For $k=4$, the auxiliary graph $H_4$ is composed of 4 connected components which correspond to the pairs of $K_5$ that share one node and the pairs of $K_5$ that share 2 nodes (see figure \ref{fig:aux4}). In terms of biconnectivity, there are six bicomponents, with the two $K_5$ that overlap in two nodes as a single bicomponent. Inside these six bicomponents there are eight 4-components, but only four of them were detected (see figure \ref{fig:subgraphs4}). This is because when we build the $H_{candidate}$ subgraphs with all nodes in each biconnected component of $H_4$ that have exactly the same core number, in the case of the two $K_5$ that overlap in two nodes, all their nodes  have the same core number (4), but their density is 0.67 and the difference in degree is 3. Thus, in order to detect them we would have to relax the clique criteria for $H_{candidate}$ too much, and even then, we would classify both $K_5$ overlaping in two nodes as a single 4-component, which is obviously wrong because they have node connectivity 2.

Note that this kind of false negative only happens when two $k$-components of the same level of connectivity and the same order overlap. If instead of two $K_5$ they were $k$-components with different order but the same connectivity, our algorithm would be able to separate them because they would have a different core number and thus they would be part of a different $H_{candidate}$ subgraph.

\section{Performance analysis}
\label{performance}

The heuristics presented here are implemented on top of NetworkX \citep{hagberg:2008}, a library for the analysis of complex networks, using the Python programming language \citep{vanrossum:1995}. We have chosen Python because it is a language with high readability and flexibility that allows you to easily apply the well know principle of writing software for people to read and, only incidentally, for machines to execute \citep{abelson:1985}. To ensure reproducibility and accessibility we have used only free software to build and run all analyses presented in this paper. 

The implementation of the heuristics presented here is not trivial; a careful implementation is needed to ensure that it has a reasonable memory footprint and that it runs in a reasonable time. Appendix C contains a detailed discussion of the implementation details and appendix D contains the python code of a simplified implementation for illustrative purposes.

\begin{figure}[h]
\centering
\subfloat[Performance of connectivity algorithms when adding nodes maintaining constant the average degree (Erdös-Rènyi) or the exponent of the power law governing the degree distribution ($\alpha=2$). Logarithmic scale.]{
\label{fig:nodes}
\includegraphics[scale=0.37]{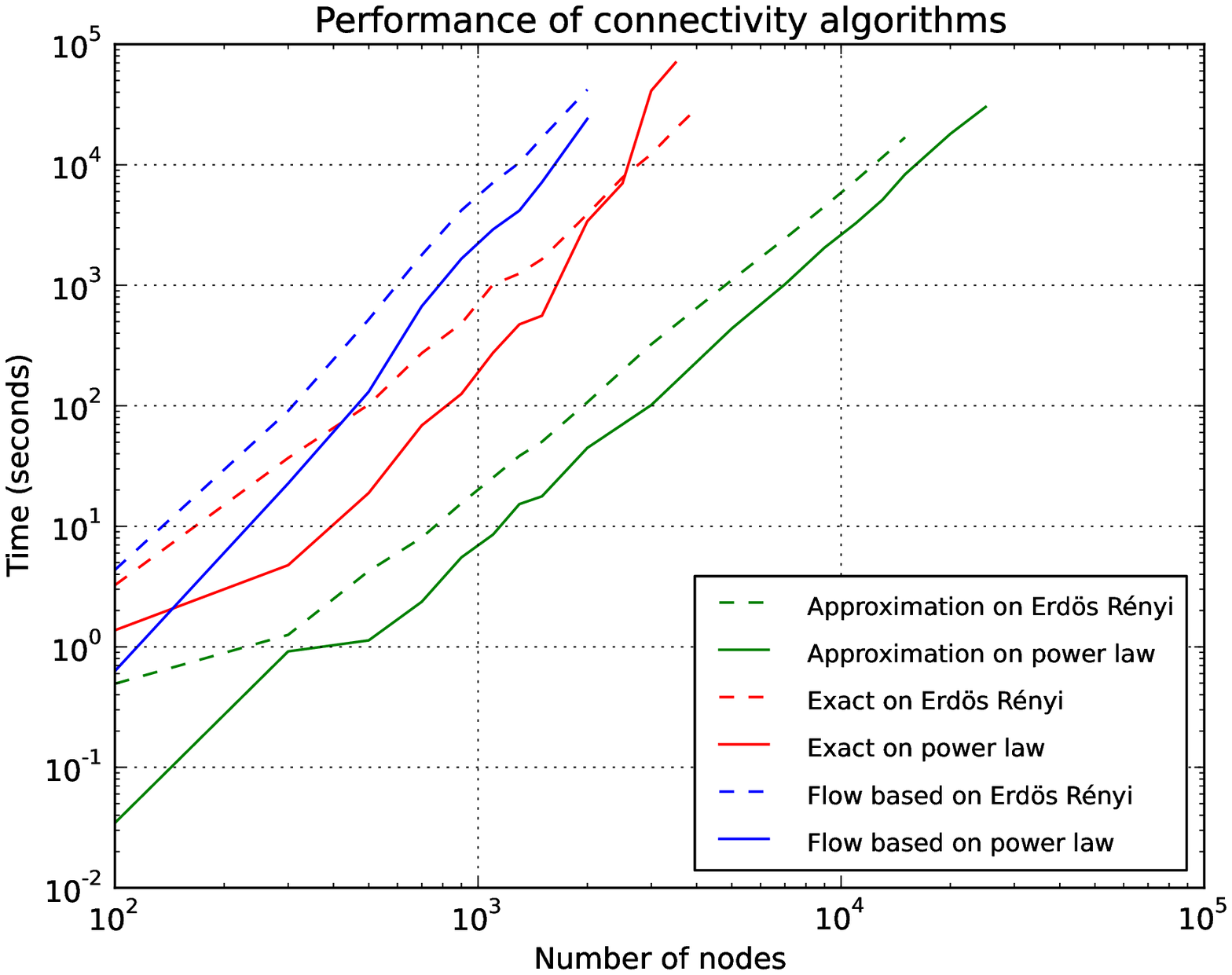}
}
\hspace{.05in}
\subfloat[Performance of the heuristics when adding edges and maintaining nodes constant (1000 nodes). Inset: performance of the exact algorithm with one order of magnitude fewer nodes (100 nodes). Both in logarithmic scale.]{
\label{fig:edges}
\includegraphics[scale=0.37]{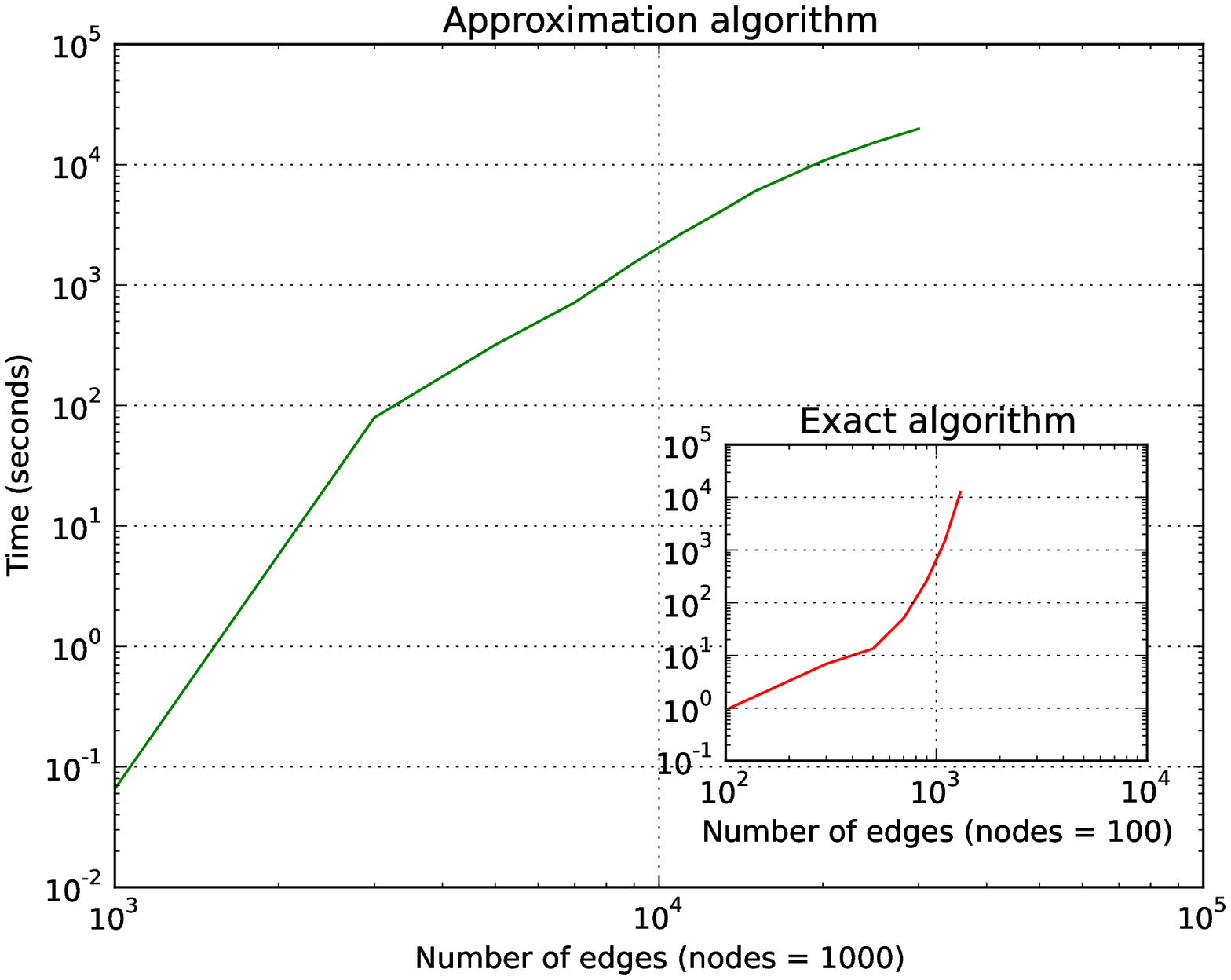}
}
%\hspace{.2in}

\begin{minipage}{\textwidth}
\caption[Loglog plots for comparing between heuristics and exact algorithms]{Loglog plots for comparing between the heuristics and the exact algorithm to compute $k$-component structure. In this comparison, the heuristics do not compute the average node connectivity, only plain node connectivity, which is what is calculated by the exact algorithm. We have also implemented the exact algorithm in order to be able to compare both algorithms using the same language and infrastructure. All figures presented here were obtained running PyPy \citep{bolz:2009}. Using the heuristics proposed in this paper, we are able to handle networks almost one order of magnitude bigger than with the exact algorithm.}
\label{fig:performance}
\end{minipage}
\end{figure}

Figure \ref{fig:performance} presents the performance of the heuristics (green) compared with two variants of the exact algorithm: the Moody \& White algorithm based on $k$-cutsets (red) and our algorithm using exact flow-based node connectivity for building the auxiliary graph. The tests were performed, on the one hand, on random graphs with fixed average degree (Erdös-Renyi model) and fixed power law exponent (Power law model) of several different orders. And, on the other hand, for graphs with a fixed number of nodes (1000 for the heuristics and 100 for the exact) where we increase the number of edges. Random networks built using the Erdös-Renyi model have a flat hierarchical structure because edges are evenly distributed across all nodes of the network. The Erdös-Renyi graphs used in this benchmark have a big tricomponent and no higher connectivity levels. Random networks built using a power law based degree distribution have a steep hierarchical structure, the networks used in the benchmark have hierarchy levels of up to 20. Both the heuristics and the exact algorithms perform better in sparse networks with a steep hierarchical structure.

As we can see in figure \ref{fig:performance} the heuristics runs in polynomial time. It is fast enough to be practically applicable to networks with a few tens of thousands of nodes and edges. This is one order of magnitude better than the exact algorithm proposed by \citet{moody:2003}, and also an order of magnitude faster than using flow-based algorithms for building the auxiliary graph. Notice that the $k$-cutset based algorithm proposed by Moody \& White (or at least our implementation) is faster than the exact flow-based local node connectivity variant of our algorithm.

The implementation that we provide in this paper only considers the exact solution for biconnected components. The heuristics presented here uses biconnectivity, but can be improved by using a triconnectivity algorithm. It would be: a) faster because there is a linear algorithm to compute triconnected components \citep{tarjan:1974,gutwenger:2001}; and, b) more accurate, because we compute the exact solution up to $k=3$. But, as far as we know, there is no publicly available implementation of triconnected components. An optimal implementation of the heuristics presented here would have to incorporate the triconnectivity algorithm to improve its accuracy and to allow it to run in reasonable time on somewhat larger networks.

\newpage

\section{Implementation details}
\label{implementation}

The implementation of the heuristics proposed here was done by the first author listed on the NetworkX python library \citep{hagberg:2008}, a Python package for the study of the structure and dynamics of complex networks. Other parts of the powerful Python \citep{vanrossum:1995} scientific computing stack \citep{scipy,ipython, hunter:2007} were also essential. The main requirement was that the whole software stack must be free software in order to avoid the black box effect of software solutions that do not release their source code. We belive that this is a necessary condition for ensuring the reproducibility of scientific research. Appendix B contains python code for the main part of the algorithm. 

The implementation of the heuristics is not trivial. There are a few questions that need to be addressed in order to obtain a performance ---both in terms of computation time and memory consumption--- that will allow for these heuristics to be applied to large networks. The authors are in-debted to Aric Hagberg and Dan Schult (developers of the NetworkX package) for their help in this implementation.

The second step of the heuristics (compute the biconnected components of the input graph and use them as a baseline for $k$-components with $k > 2$) is faster than using the logic of the heuristics for $k=2$. Biconnected components computation runs in linear time in respect to the number of nodes and edges \citep{tarjan:1972}. Besides in large networks, bicomponents are formed by an important part of the nodes of the network. Thus if we use the approximation logic to compute them, the memory footprint for large networks is too large to be practical. The implementation provided with this paper only computes the exact solution for bicomponents but there is also a linear algorithm to compute triconnected components \citep{tarjan:1974,gutwenger:2001}. The heuristics would be even faster if we applied the approach used for bicomponents to that of tricomponents. But the implementation of triconnectivity is quite challenging and, to our knowledge, there is no implementation of triconnected components in free network analysis software packages.

The auxiliary graph $H$ is usually very dense in real world networks because a large part of nodes that are in a biconnected part of a $k$-core are actually part of a $k$-component. The memory footprint of creating this dense auxiliary graph prevents a naive implementation of the heuristics in order to be practical for large networks. Our solution for this problem is to use a complement graph data structure that only stores information on the edges that are \emph{not} present in the actual auxiliary graph. When applying algorithms to this complement graph data structure, it behaves as if it were the dense version. This is the only way to have a memory footprint that will allow for the application of the heuristics presented in this paper to large networks.

\newpage

\section{Python code}
\label{code}

This is a simplified implementation of the heuristics for illustrative purposes. A fully functional version of NetworkX package with all the code necessary to run the heuristics is available from the authors upon request. 

%can be found at \href{https://bitbucket.org/jtorrents/networkx\_sc}{https://bitbucket.org/jtorrents/networkx\_sc}.

\begin{footnotesize}
\begin{lstlisting}
# Standard python libraries
import itertools
import collections
# NetworkX library for network analysis
import networkx
# White and Newman node connectivity approximation
# Code in https://networkx.lanl.gov/trac/ticket/538
from connectivity_approx import vertex_connectivity_approx
# AntiGraph data structure
# code in https://networkx.lanl.gov/trac/ticket/608
import antigraph

def k_components(G, average=True, exact=False, min_density=0.95):
    def _update_results(k, avg_k, components):
        # Auxiliary function to update results data structures
        # Code not shown
    if exact:  # Use flow based exact algorithm
        node_connectivity = nx.local_node_connectivity
    else: # Use White and Newman (2001) approximation algoritm
        node_connectivity = local_node_connectivity 
    ## Data structures to return results
    # Dictionary with connectivity level (k) as keys and a list of
    # sets of nodes that form a k-component as values
    k_components = collections.defaultdict(list)
    # Dictionary with nodes as keys and maximum k of the deepest 
    # k-component in which they are embedded
    k_number = dict( ((n,(0,0)) for n in G) )
    # dict to store node independent paths
    nip = {} 
    #################
    # Exact solution for k = 1
    components = networkx.connected_components(G)
    _update_results(1, 1, components)
    # Bicomponents as a base to check for higher order k-components
    bicomponents = networkx.biconnected_components(G)
    _update_results(2, 2, bicomponents)
    # There is no k-component of k > maximum core number
    # \kappa(G) <= \lambda(G) <= \delta(G)
    g_cnum = core_number(G)
    max_core = max(g_cnum.values())
    for k in range(3, max_core + 1):
        C = k_core(G, k, core_number=g_cnum)
        for nodes in biconnected_components(C):
            # Build a subgraph SG induced by the nodes that are part of
            # each biconnected component of the k-core subgraph C.
            if len(nodes) < k:
                continue
            SG = G.subgraph(nodes)
            # Build auxiliary graph
            H = AntiGraph()
            H.add_nodes_from(SG.nodes_iter())
            for u,v in combinations(SG, 2):
                K = node_connectivity(SG, u, v)
                nip[(u,v)] = K
                if k > K:
                    H.add_edge(u,v)
            for h_nodes in biconnected_components(H):
                if len(h_nodes) <= k:
                    continue
                HS = H.subgraph(h_nodes)
                h_cnum = core_number(HS)
                first = True
                for c_value in sorted(set(h_cnum.values()),reverse=True):
                    cands = set(n for n, cnum in h_cnum.items() if cnum == c_value)
                    # Skip checking for overlap for the highest core value
                    if first:
                        overlap = False
                        first = False
                    else:
                        overlap = set.intersection(*[
                                    set(x for x in HS[n] if x not in cands)
                                    for n in cands])
                    if overlap and len(overlap) < k:
                        Hc = HS.subgraph(cands | overlap)
                    else:
                        Hc = HS.subgraph(cands)
                    if len(Hc) <= k:
                        continue
                    hc_core = core_number(Hc)
                    if _same(hc_core) and density(Hc) == 1.0:
                        Gc = k_core(SG.subgraph(Hc), k)
                    else:
                        while Hc:
                            Gc = k_core(SG.subgraph(Hc), k)
                            Hc = HS.subgraph(Gc)
                            if not Hc:
                                continue
                            hc_core = core_number(Hc)
                            if _same(hc_core) and density(Hc) >= min_density:
                                break
                            hc_deg = Hc.degree()
                            min_deg = min(hc_deg.values())
                            remove = [n for n, d in hc_deg.items() if d == min_deg]
                            Hc.remove_nodes_from(remove)
                    if not Hc or len(Gc) <= k:
                        continue
                    for k_component in biconnected_components(Gc):
                        if len(k_component) <= k:
                            continue
                        Gk = k_core(SG.subgraph(k_component), k)
                        num = 0.0
                        den = 0.0
                        for u,v in combinations(Gk, 2):
                            den += 1
                            num += (nip[(u,v)] if (u,v) in nip
                                        else nip[(v,u)])
                        _update_results(k, [Gk.nodes()], (num/den))
    return k_components, k_number
\end{lstlisting}
\end{footnotesize}

\newpage

\section{Accuracy and limitations of the heuristics}
\label{accuracy}

\begin{figure}[p]
\centering
\subfloat[Bipartite network formed by developers and packages over 2 years of collaboration (from 2007 to 2009) on the release codenamed Lenny of the Debian operating system]{
\label{fig:lenny2m}
\includegraphics[scale=0.35]{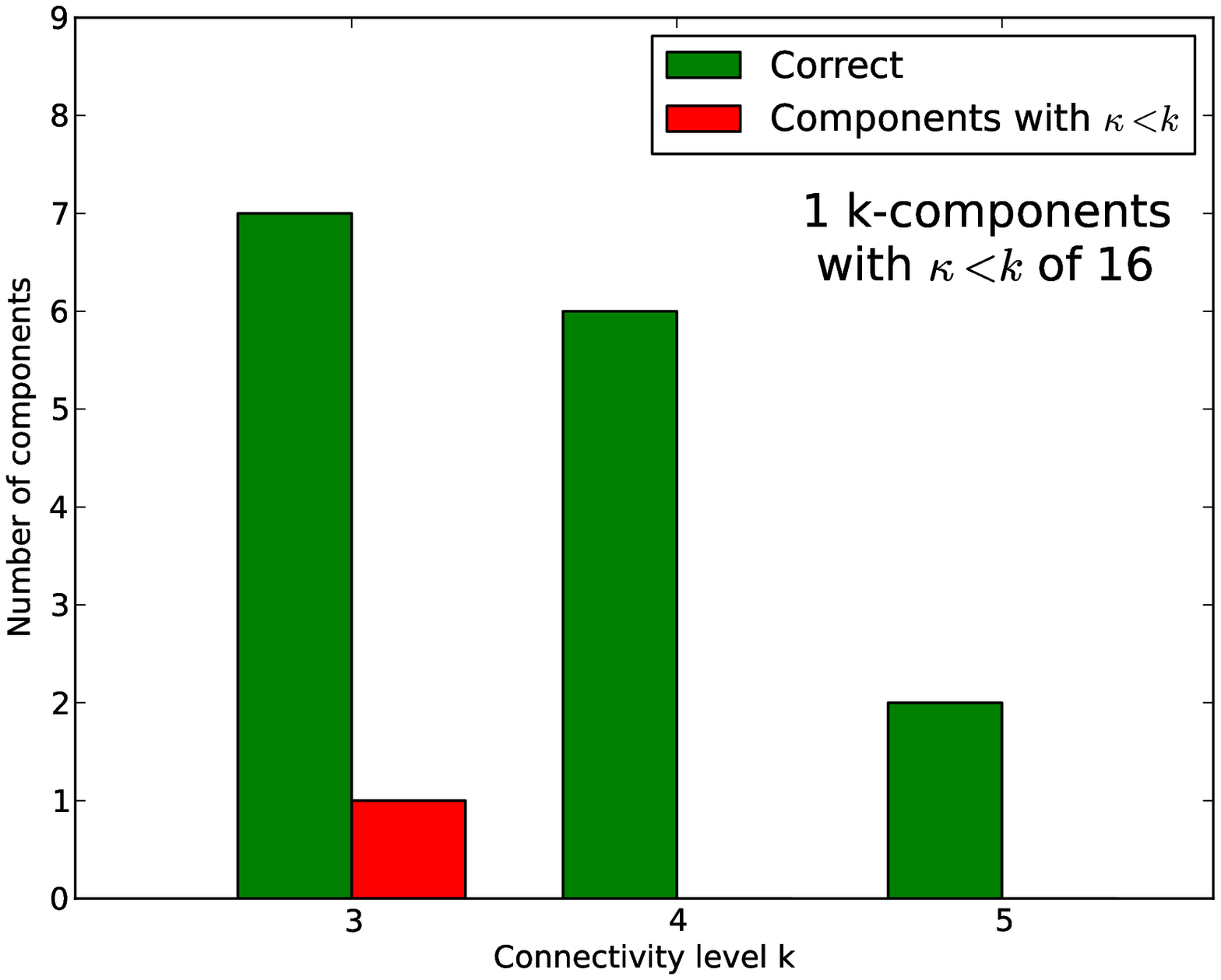}
}
\hspace{.05in}
\subfloat[Unipartite network formed by developers over 2 years of collaboration (from 2007 to 2009) on the release codenamed Lenny of the Debian operating system]{
\label{fig:lenny1m}
\includegraphics[scale=0.27]{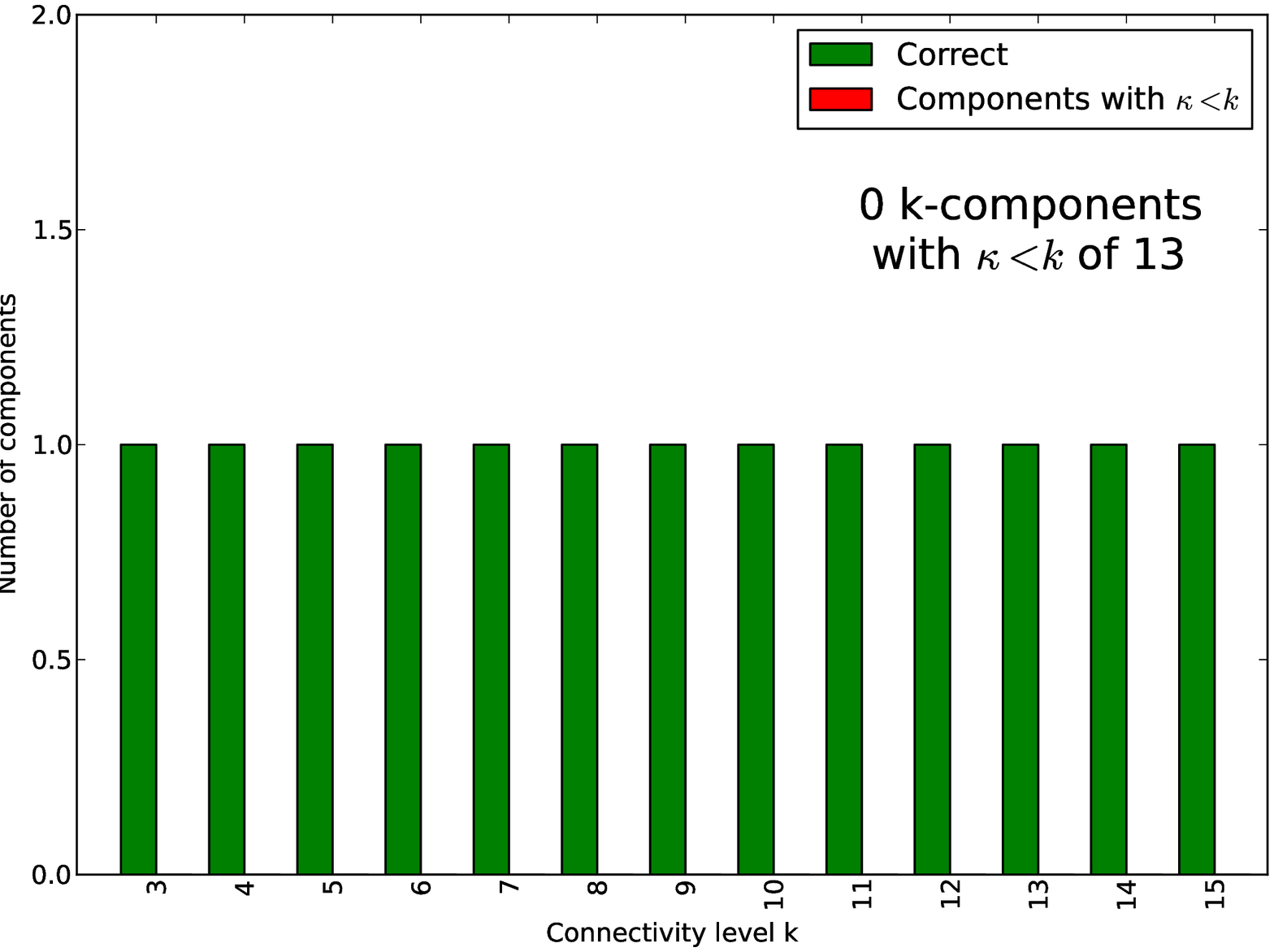}
}
\hspace{.01in}
\subfloat[Bipartite network formed by scientists and preprints during a five-year period (2006-2010) in the high energy physics (theory) section of arXiv.org]{
\label{fig:hep_th_2m}
\includegraphics[scale=0.35]{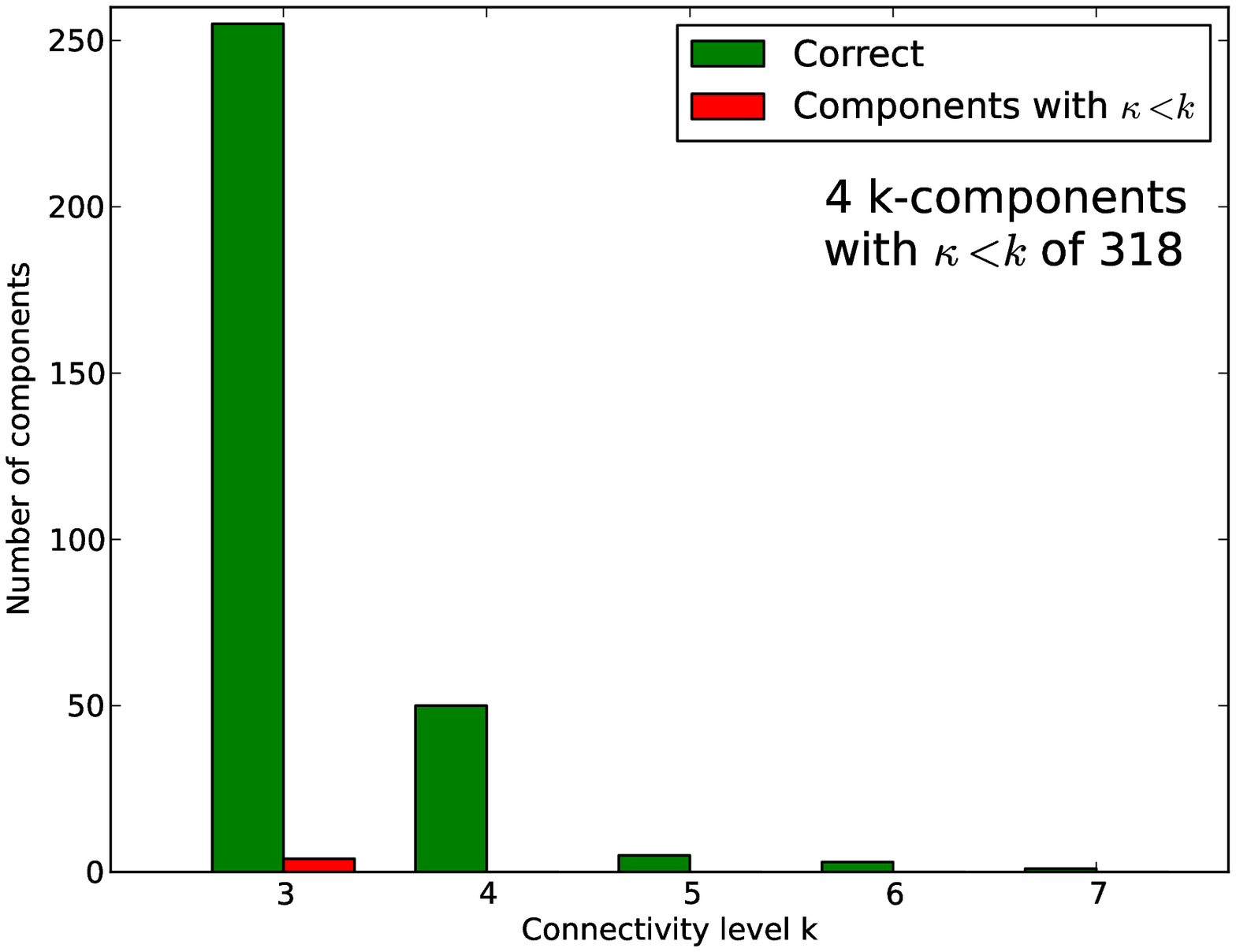}
}
\hspace{.05in}
\subfloat[Unipartite network formed by scientists during a five-year period (2006-2010) in the high energy physics (theory) section of arXiv.org]{
\label{fig:hep_th_1m}
\includegraphics[scale=0.30]{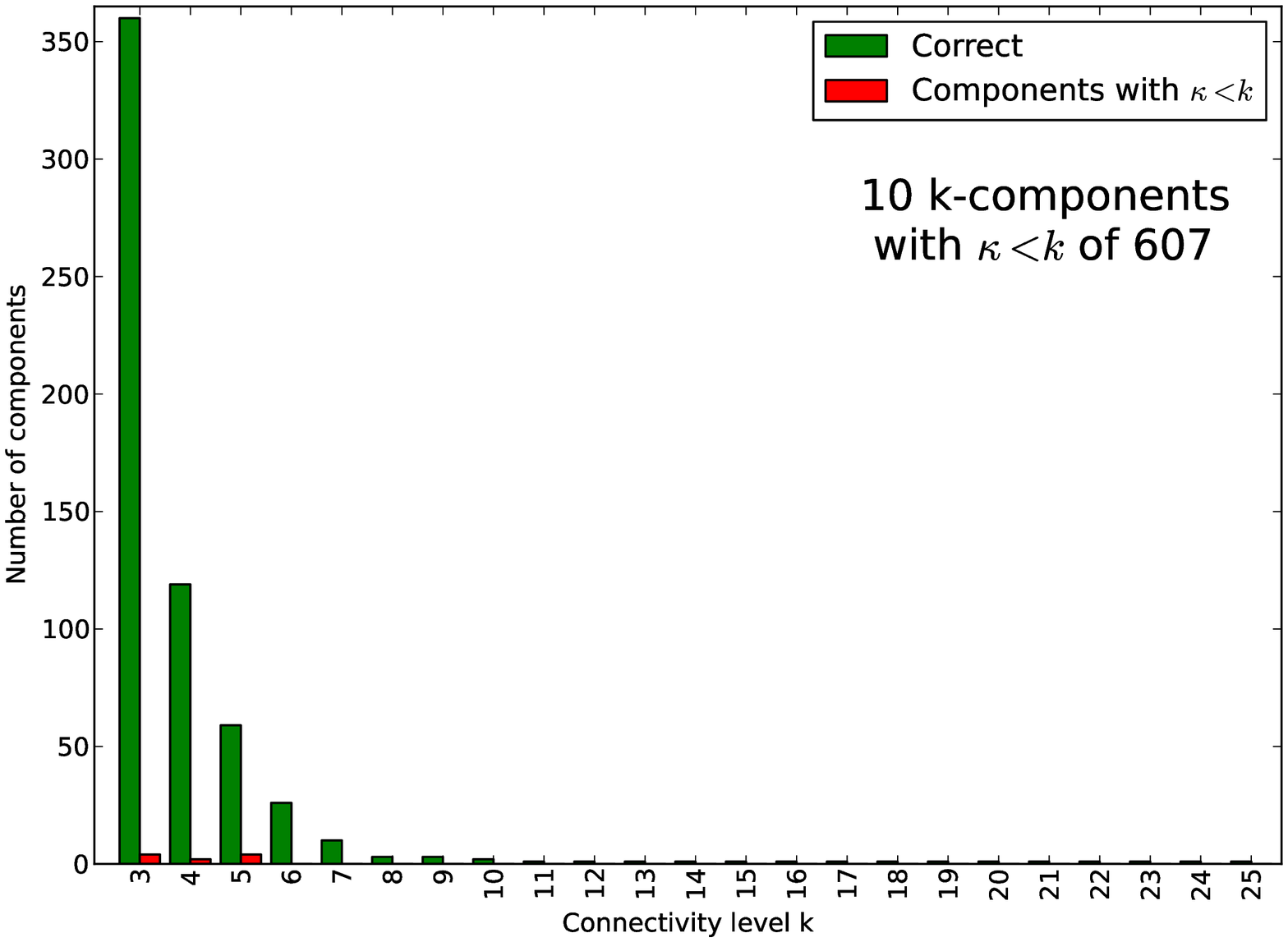}
}
\hspace{.01in}
\subfloat[Bipartite network formed by scientists and preprints during a five-year period (2006-2010) in the nuclear physics (theory) section of arXiv.org]{
\label{fig:nucl_th_2m}
\includegraphics[scale=0.35]{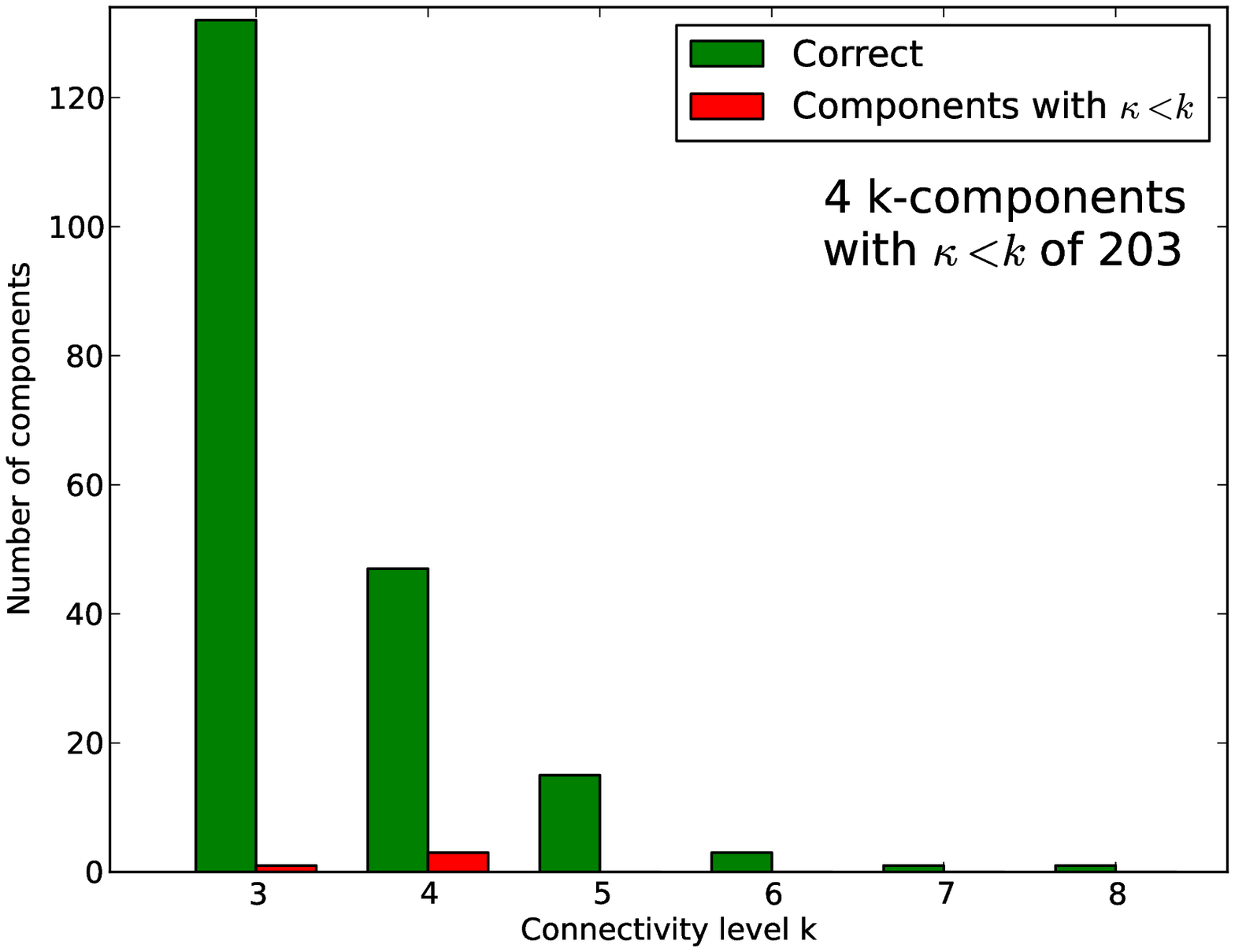}
}
\hspace{.05in}
\subfloat[Unipartite network formed by scientists during a five-year period (2006-2010) in the nuclear theory section of arXiv.org]{
\label{fig:nucl_th_1m}
\includegraphics[scale=0.27]{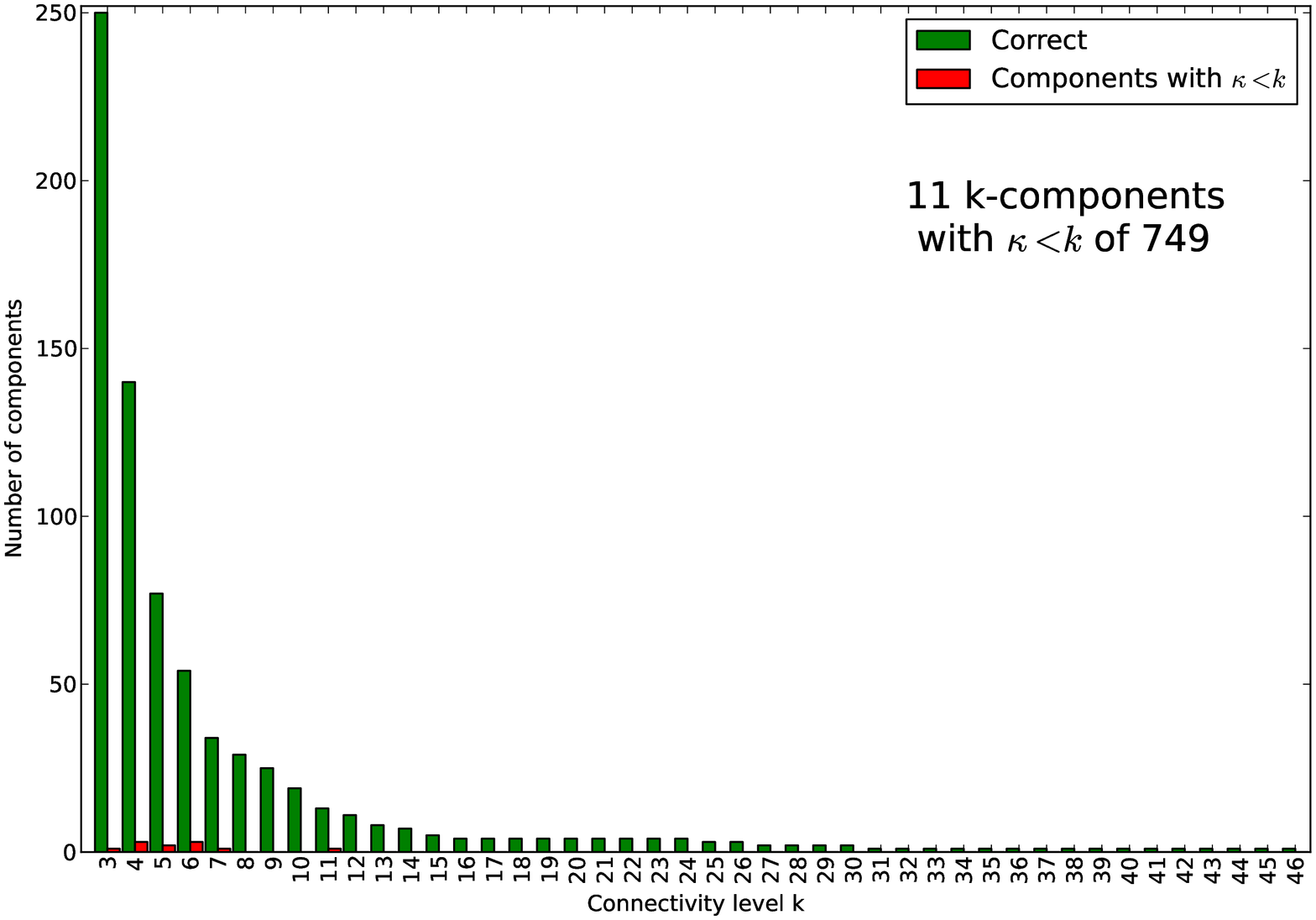}
}
\caption{Accuracy barplots. Green bars are $k$-components with node connectivity $\ge k$ and red bars represent $k$-components with node connectivity $< k$.}
\label{fig:accuracy}
\end{figure}

Figure \ref{fig:accuracy} shows the accuracy of connectivity structure detected by the heuristics for all empirical networks. In the subfigures, green bars are $k$-components with node connectivity $\ge k$ and red bars represent $k$-components with node connectivity $< k$.  Note that, once we have an approximate structure of $k$-components, we can check ---in a reasonable time frame--- if the resulting $k$-components actually have node connectivity $k$ using flow based connectivity algorithms \citep[chapter 7]{brandes:2005}. For the candidate $k$-components that turned out to have node connectivity lower than $k$, we used the exact algorithm proposed by \citet{moody:2003} to find out the order and size of the actual $k$-components inside the candidate $k$-component detected using our heuristics.

The output of our heuristics is an approximation to $k$-components based on computing extra-cohesive blocks for each biconnected component of all core levels of the network. Recall that in $k$-components all $k$ node independent paths go through nodes that belong to the $k$-component, but in extra-cohesive blocks some of the node independent paths may go through external nodes. Thus, there is no guarantee that the extra-cohesive blocks, even those that also form a $k$-core subgraph in $G$, have node connectivity $\kappa = k$. This is a source of false positives for the approximation of the $k$-component structure of a network. However, the results shown in figure \ref{fig:accuracy} suggest that the heuristics yield a good approximation for the actual ---$k$-component based--- cohesion structure of empirical networks.

If we consider all components of all sizes, as in figure \ref{fig:accuracy}, only a few of the extra-cohesive blocks detected by the heuristics have node connectivity of less than $k$, ranging from 6.5\% (a single component) in the case of Debian to 1.2\% of the components in the case of two-mode Nuclear Theory network. However, the extra-cohesive blocks that do not have the sufficient connectivity to be considered a $k$-component are, in the empirical networks analyzed, big components of levels \{3,4\}. This is because, in such big- and low-level components, a few node independent paths going through nodes that are part of the biconnected component of a $k$-core but not part of the $k$-component can yield false positives by including nodes that shouldn't be part of the $k$-component.

However, these false positives are actually part of an extra-cohesive block, which maintains most of those properties ---in terms of robustness, hierarchy and overlap--- which make $k$-component such a good measure of structural cohesion. This relaxed definition of connectivity might be sufficient in many cases; for instance, if we are interested in comparing the structural cohesion of a large network with a suitable null model, we may not need the exact $k$-component structure because we can meaningfully compare the relaxed connectivity structure of the actual network with its random counterparts. However, imagine we are interested in the exact $k$-component structure of a particular network because, say, we want to statistically analyze the impact of the connectivity level with the performance of different actors in a network. In this case, we would need to apply some cutting procedure on the extra-cohesive blocks that actually have a node connectivity of less than $k$.

It is more difficult to assess the impact of false negatives ---that is, nodes that should be part of a $k$-component but are excluded--- because computing exact $k$-components for big networks is not practical, and thus we cannot compare. False negatives are derived from the underestimation of local node connectivity of the \citet{white:2001b} algorithm, which provides a strict lower bound for the local node connectivity. Thus, by using it we can miss an edge in the auxiliary graph $H$ that should be there. Therefore, a node belonging to a $k$-component could be excluded by the algorithm. Recall that in order to address this problem, we relaxed the clique criteria by setting a density threshold of 0.95 in $H_{candidate}$. Whilst this value has worked well in our analysis but careful experimentation should be performed to set this parameter in other types of networks.

\newpage

% Bibliografia
\bibliographystyle{chicago}
\bibliography{structural_cohesion}

\end{document}